\def\apj{{\rm ApJ}}
\def\pasp{{\rm PASP}}
\def\pasj{{\rm PASJ}}
\def\mnras{{\rm MNRAS}}

\def\apjs{{\rm ApJS}}
\def\aap{{\rm A\&A}}
\def\araa{{\rm ARA\&A}}
\def\aaps{{\rm A\&AS}}

\def\apss{{\rm APS Conf. Ser.}}
\def\apjl{{\rm ApJL}}
\def\aj{{\rm AJ}}

\documentclass{aa}
\usepackage{graphicx}
\usepackage{latexsym}
\usepackage{amsmath}
\usepackage{amssymb}

\newcounter{qub}
\begin{document}

\title{X-RAY NATURE OF THE LINER NUCLEAR SOURCES}

\author{
O. Gonz\'alez-Mart\'in \inst{1},
J. Masegosa \inst{1},
I. M\'arquez \inst{1},
M. A. Guerrero \inst{1}
 \and D. Dultzin-Hacyan \inst{2}
}

\offprints{O.~Gonz\'alez-Mart\'in \email{omaira@iaa.es}}

\institute{
Instituto de Astrof\'isica de Andaluc\'ia, CSIC, Apartado
 Postal. 3004, 18080, Granada, Spain \and Instituto de Astronom\'ia,
 UNAM, Apartado Postal 70-264, 04510, M\'exico D.F. , M\'exico}
\date{} \abstract{We report the results from an homogeneous analysis
of the X-ray (\emph{Chandra} ACIS) data available for a sample of 51
LINER galaxies selected from the catalogue by Carrillo et al. (1999)
and representative of the population of bright LINER sources. The
nuclear X-ray morphology has been classified attending to their
nuclear compactness in the hard band (4.5--8.0~keV) into 2 categories:
Active Galactic Nuclei (AGN) candidates (with a clearly identified
unresolved nuclear source) and Starburst (SB) candidates (without a
clear nuclear source). 60\% of the total sample are classified as
AGNs, with a median luminosity of $\rm{L_{X}(2-10~keV)=2.5\times
10^{40}~erg~s^{-1}}$, which is an order of magnitude higher than that
for SB-like nuclei. The spectral fitting allows to conclude that most
of the objects need a non-negligible power-law contribution. When no
spectral fitting can be performed (data with low signal-to-noise
ratio), the Color-Color diagrams allow us to roughly estimate physical
parameters such as column density, temperature of the thermal model or
spectral index for a power-law and therefore to better constrain the
origin of the X-ray emission.  All together the X-ray morphology, the
spectra and the Color-Color diagrams allow us to
conclude that a high percentage of LINER galaxies, at least $\approx$
60\%, could host AGN nuclei, although contributions from High Mass
X-ray Binaries or Ultra-luminous X-ray sources cannot be ruled out for
some galaxies.
\keywords{galaxies, AGN, LINER, X-ray, Chandra}}
\authorrunning{O. Gonz\'alez-Mart\'in et al.}

\titlerunning{X-Ray Nature of LINERs}

\maketitle

\section{Introduction}
Active Galactic Nuclei (AGN) produce enormous luminosities in
extremely compact volumes. Large luminosity variations on time scales
from years to hours are common (e.g. Leighly 1999). The combination of
high luminosity and short variability time scales implies that the
power of AGN is produced by phenomena more efficient in terms of
energy release per unit mass than ordinary stellar processes (Fabian
1979).

A quantitative definition of what constitutes an active galaxy is
perhaps not very useful, since galaxies showing low-level activity
(e.g. Heckman 1980, Stauffer 1982, Hawley \& Philips 1980) may be in
a pre- or post- eruptive stage, and so may yield important clues into
the origin and evolution of nuclear activity. The question of whether
similar unification ideas can also apply to low-luminosity AGNs
(LLAGNs), which make up the vast majority of the AGN population, has
been explored (see Barth 2002). In this sense, LLAGNs might constitute
a perfect laboratory to investigate the connection between galaxies in
which the central black holes are active and those in which they are
quiescent. However, such AGNs may be difficult to identify because of
extinction (e.g. Keel 1980, Lawrence \& Elvis 1982)
or contamination by star forming processes in circunnuclear region
(e.g. V\'eron et al. 1981). The number of weak AGNs increases every
time deep searches are made. Heckman (1980, see also Heckman et al. 
1980 and Ho et al. 1997) has shown that one third of a
complete sample of 'normal' galaxies exhibit signs of nuclear
activity. 


LINERs (Low-Ionization Nuclear Emission-line Regions) were originally
defined as a subclass of these LLAGNs by Heckman (1980) and are
characterized by optical spectra dominated by emission lines of
moderate intensities arising from gas in lower ionization states than
classical AGNs. LINERs where defined as galaxies whose spectra satisfy
$\rm{[OII]\lambda 3727/[OIII]\lambda 5007 \ge 1}$ and $\rm{[OI]\lambda
6300/[OIII]\lambda 5007 \ge 1/3}$ (Heckman 1980). These LINERs
typically are less luminous than powerful Seyferts and
QSOs. It is still unclear whether all LINERs are essential AGNs
at all, but if LINERs represent the low-luminosity end of the AGN
phenomenon, they are the nearest and most common examples, and their
study is genuine to understanding AGN demographics and evolution. One
fundamental question that needs to be addressed is whether the nuclear
emission of these galaxies results from starbursts or accretion onto
super-massive black holes (SMBHs).

The low luminosity of these nuclear sources makes them difficult
targets for observational studies, even in very nearby galaxies. The
origin of the optical narrow emission lines of LINERs has long been a
source of controversy because the optical line ratios can be
reproduced reasonably well by models based on a variety of different
physical mechanisms, including shock heating (Fosbury et al. 1978;
Dopita \& Sutherland 1996), photo-ionization by a non stellar continuum
(Ferland \& Netzer 1983; Halpern \& Steiner 1983), photo-ionization by
young starburst containing Wolf-Rayet stars (Filippenko \& Halpern
1984; Terlevich \& Melnick 1985; Barth \& Shields 2000), or
photo-ionization by hot stars (Filippenko \& Terlevich 1992; Shields
1992).

One would expect that information coming from different spectral ranges
could help disentangle and eventually taxonomies the LINER
family. However, the study of data at different wavelengths has provoked
even more discussion.

UV imaging surveys by Barth et al. (1998) and Maoz et al. (2005) 
(also Pogge et al. 2000) found
nuclear UV emission in $\sim$25\% of the LINERs that were
observed. About half of them appear point-like at the resolution of HST and
thus are good candidates for being genuine LLAGNs with non-stellar
continua. Barth et al. (1998) showed that the low UV detection rate is
primarily due to dust obscuration of the nuclei. Thus, the majority of
LINERs probably have UV sources in their nuclei (which could be
either AGNs or young star clusters), but in most cases the UV sources
lie behind large amount of dust to render it visible. Therefore, the foreground 
dust plays an important role in blocking our view of the central engines.


 Since the nuclei of LINERs may be very heavily obscured, observations
 in UV, optical, near-infrared, and even the far-infrared may not
 penetrate through out the dust to reach the nucleus. The most important
 recent data constraining the nature of LINER nuclei have come from
 radio and X-rays surveys because in these spectral regions it is
 possible to detect central engines that are completely obscured in
 the optical and UV. In a VLA survey, Nagar et al. (2000) found that
 64\% of LINER 1 and 36\% of LINER 2 have compact radio cores (Nagar et al.
 2005). The objects bright enough for VLBI observations at 5GHz were studied by
 Falcke et al. (2000, see also Filho et al. 2004); all showed compact, 
 high-brightness-temperature cores, suggesting that an AGN rather than a 
 starburst is responsible for the radio emission. Moreover, the core radio 
 fluxes have been found to be variable by a factor up to a few in about half 
 of the $\sim$10 LINERs observed multiple times over 3 years (Nagar et
 al. 2002). A radio survey for 1.3 cm water mega-maser emission, an
 indicator of dense circunnuclear molecular gas, detected LINER nuclei
 at the same rate as type 2 Seyfert nuclei (Braatz et al. 1997). Such
 mega-maser emission is seen only in AGNs. Some LINERs have
 indications of a Seyfert-like ionization cone oriented along their
 radio axis (Pogge et al. 2000).

X-ray observations provide another direct probe of the central
engines.  Pure starburst galaxies, at low redshift, do not exhibit
unresolved hard X-ray (2.0-8.0~keV) nuclei. On the contrary,
starbursts such as M82, have extended hard X-ray emission from both
diffuse gas and unresolved X-ray binaries (Griffiths et al. 2000). The detection
of a hard X-ray continuum, as well as Fe K emission (Iyomoto et
al. 1996; Ishisaki et al. 1996; Terashima et al. 1998, 1999, 2000;
Roberts, Warwick and Ohashi 1999; Jimenez-Bailon et al. 2005; Streblyanska et
al. 2005) are indicators of AGN
activity. Only a reduced number of X-ray observations of LINER 2s have
been performed so far. X-ray observations with \emph{Einstein} and
\emph{ROSAT} were limited to soft energies, where heavily obscured AGNs are
difficult to detect. \emph{ROSAT HRI} images showed compact soft X-ray ($\rm{\le 2~keV}$)
emission in 70\% of LINERs and Seyfert galaxies (Roberts \& Warwick
2000). Nevertheless, the lack of spectral information, low spatial 
resolution and inadequate bandpass of these observations cannot 
distinguish the thermal emission of the host galaxy from the emission 
from the AGN. These problems are overcame by \emph{Chandra}, whose 
spatial resolution 10 times superior to that of the \emph{ROSAT HRI} 
allows us to resolve the emission on lower physical sizes.

In this paper we present the homogeneous analysis of a sample of
LINERs observed by \emph{Chandra} and examine the probable ionization
mechanism in LINERs. The paper is organized as follows. In
Section 2 we summarize the \emph{Chandra} and HST observations and
describe the galaxy sample. Image and spectral reduction and analysis
of X-ray data and HST imaging are reported in Section 3. We
discuss the origin of the X-ray emission in our sample in Section
4. Finally, a summary of our findings is present in Section 5.

\section{The Sample and the Data}

The starting list for the sample selection has been the
Multi-wavelength catalogue of LINERs (MCL) compiled by Carrillo et
al. (1999).  MCL includes most of the LINER galaxies known until 1999,
providing information on broad band and monochromatic emission from
radio frequencies to X-rays for 476 objects classified as LINERs. The
initial galaxy sample was constructed by selecting in the \emph{Chandra}
archive all the galaxies in MCL with Advanced CCD Imaging Spectrometer
(ACIS\footnote{Comprising two back-illuminated CCD chips and eight
front-illuminated CCD chips of 1024 pixels square, with a plate scale
of 0."492 $\rm{pixel^{-1}}$.})  observations already public in
November 2004, what yielded a set of 137 out of the 476 galaxies.

The optical classification was reanalyzed (by using the line ratios
diagrams by Veilleux \& Osterbrock 1987), and 15 objects were
eliminated from the sample due to misclassification in MCL; 5 of them
appear to be Starburst systems (NGC\,1808, NGC\,3077, ESO\,148-IG002N,
ESO\,148-IG002S and NGC\,253), 6 are Seyfert-like galaxies (NGC\,4258,
MRK\,0266SW, 3C452, NGC\,4565, NGC\,4501 and NGC\,3079) and 4 are
transition objects (NGC\,0224, NGC\,0404, AN\,0248+43B and
NGC\,4303).  For the selected sample we have noticed that data
with exposure times lower than 10~ksec had less than 25 counts in the
0.5-10.0~keV energy range.  Therefore, only data with larger exposure
times have been taken into account. The final sample, with high
quality data and optical re-identification as LINER nuclei, amounts to
51 objects.  Almost all objects (except NGC\,3607, NGC\,3608,
NGC\,3690B, NGC\,4636, NGC\,5746 and NGC\,6251) have been observed
with ACIS-S mode.  The 51 galaxies were observed between August 2000
and April 2004. A list of the objects, including the details of their
observations, is provided in Table
\ref{tab:obsdata} in which Name (Col. 1), X-ray position in right
ascension and declination (Cols. 2 and 3), X-ray radii selected for
the nuclear sources (Col. 4) and offset with respect to the 2MASS
coordinates (Col. 5), \emph{Chandra} Observational Identifier (Col. 6)
and exposure time after removal of background flares (Col. 7) are
given.

The data provided in Table \ref{tab:catdata} have been extracted
from Carrillo et al. (1999) and include properties of the host
galaxies such as: Source name (Col. 1), Redshift z (Col. 2), distances
(Col. 3), spatial scale at the distance of the galaxy
(Col. 4), source radii of the selected X-ray nuclear source regions (Col. 5),
B magnitude (Col. 6), E(B-V) (Col. 7), and Morphological Type
(Col. 8). The codes for the assumed distances correspond to those extracted 
from (a) Ferrarese et al. 2000;
(b) assuming a cosmology with $\rm{H_{o}}=75 Km~s^{-1}~Mpc^{-1}$ and
$\rm{q_{o}}=0$; (c) Tonry et al. 2001; (d) Tully 1998; 
and (e) Karachentsev \& Drodovsky 1998.

In Fig. \ref{fig:hist1}, from top to bottom, the normalized redshifts,
morphological types and absolute and apparent magnitudes distributions
are shown for the MCL catalogue (empty histogram) and the X-ray sample
(filled histogram). On the MCL catalogue, the z distribution shows
that most of the LINERs are hosted in nearby galaxies
(Fig. \ref{fig:hist1}a); the Hubble type histogram shows that host
galaxies of LINERs are mainly normal spirals (Fig. \ref{fig:hist1}b)
with a median B magnitude of $\rm{M_{B}=-20.0\pm 1.5}$
(Fig. \ref{fig:hist1}c). It has to be noticed that whereas the
redshift and absolute magnitude distributions are very similar to
those of the total sample, the X-ray selected sample resides mostly in
earliest Hubble type galaxies. This bias might be produced by the way
in which these galaxies have been selected for observations: they are
part of guaranteed and open-time programs with differing scientific
goals.
In Fig. \ref{fig:hist1}d it can be seen the bias produced by the apparent
magnitude selection. Most of the galaxies in the X-ray sample come 
from the bright LINER sample cataloged by Ho et al. (1997), but for some particular
peculiar cases that where observed because of their interest:
 NGC\,6240, UGC\,08696 and UGC\,05101 belong to the class of Ultra-luminous
Infrared Galaxies (ULIRGs); CGCG\,162-010 is the central cluster galaxy
in Abell\,1795, and NGC\,0833 (Arp\,318B) lies in a Hickson compact
group, HCG\,16.  Therefore we believe that this sample can be
considered representative only of the bright galaxy population but not
for all the LINER population. It has to be noticed, for instance, that
the sample does not include the strong IR emitters, that seem to be a
large percentage of all known LINERs (Veilleux et al. 1999; Masegosa \&
M\'arquez 2003); in fact, the LINER galaxies in our X-ray sample 
with far IR data from the IRAS Point Source Catalogue appear to be 
rather faint IR emitters with an average IR luminosity of 10$^{10}erg~s^{-1}$.


\begin{figure}
  \caption{(a) Redshift, (b) morphological types (from the RC3
catalog: t$<$0 are for ellipticals, t=0 for S0, t=1 for Sa, t=3 for
Sb, t=5 for Sc, t=7 for Sd, and t$>$8 for Irregulars) (c) absolute
magnitudes and (d) apparent magnitudes distribution for the total
sample of LINERs in MCL (empty histogram) and for our X-Ray sample
(full histogram), normalized to the number of objects in each
sample.}\label{fig:hist1}
\end{figure}

Together with \emph{Chandra} X-ray data, we will make use of the
high-resolution, optical information provided by HST imaging for our
sample galaxies.  45 out of the total 51 galaxies have been
observed with WFPC2 in several different programs, so with different
filters and exposure times. The observations in the red broad filter
F814W have been selected (31 galaxies), but observations in other
broad filters (mainly F606W and F702W) have been used otherwise. 
The summary of the selected data is given in Cols. 8, 9 and 10 in
Table
\ref{tab:obsdata}, including filter, proposal identifier and exposure 
time of the archival data. In Section 3.2 the analysis of HST data is
described.

\section{Data Reduction and Analysis}
\subsection{X-ray Data}
Level 2 event data from ACIS instrument have been extracted
from \emph{Chandra} archive. The data products were analyzed in an uniform, 
self-consistent manner using \emph{CXC Chandra Interactive Analysis of Observations}
(CIAO\footnote{See http://asc.harvard.edu/ciao}) software version
3.1. The spectral analysis was performed with \emph{XSPEC}\footnote{See
http://cxc.heasarc.gsfc.nasa.gov/docs/xanadu/xspec/} (version
11.3.2). 
Background "flares" (periods of enhanced count rate) can
seriously affect the scientific value of an observation, increasing the
count rate by a factor of up to 100. These "flares" are due to
low-energy protons interacting with the detector. Such flares have
been observed anywhere in the orbit, including near the apogee,
consequently there are not due to Van Allen belt effect.
Images could be dominated by the background if time intervals affected 
by flares are not excluded. The exposure time was therefore processed to
exclude background flares, using the task {\sc lc\_clean.sl}\footnote{see
http://cxc.harvard.edu/ciao/download/scripts/} in source-free sky
regions of the same observation.
For all observations (except NGC\,4486, excluded in the spectral
analysis)
the nuclear counts were insufficient for photon pile-up to be
significant.

\subsubsection{Spectral Analysis}

To discriminate what emission mechanisms are involved in these objects
and to estimate the X-ray luminosity 
a careful analysis of the spectra, based on 
model fittings, has been performed.
X-ray luminosities in the hard band (2.0--10.0 keV) can
be otherwise estimated following Ho et al. (2001), who assume a power
law with a spectral index of 1.8 for the SED. Since most of our
objects show a clear compact nuclear source, suggestive of AGN nature, 
large departures from a power-law index 1.8 are not expected (see Terashima
1999). Nevertheless, this approach has been shown to be too simplistic
in some cases (i.e. NGC\,3077, Ott et al. 2003), hence our attempt to
calculate luminosities {\sl via} the spectral fitting.

To extract the nuclear spectra, we first determined the position
of the nuclear sources as cataloged by near-IR observations from the
Two Micron All Sky Survey (2MASS) (see Fig. \ref{fig:contour}).
NGC\,4636 and NGC\,4676B have not been found in the near-IR catalog, so their
positions from NED have been taken as reference.

Nuclear spectra were extracted from a circular region centred in the object 
using regions defined to include as
many of the source photons as possible, but at the same time
minimizing contamination from nearby sources and
background. 
In order to determine appropriate source extraction, the radius of
each source aperture on the detector was estimated as follows: (1)
4--6 pixels for a single source, (2) 3--4 pixels for objects with few
knots close to the nucleus to exclude nearby sources, and (3)
$\rm{\ge}$~6 pixels for sources dominated by diffuse emission, since a
good S/N is required for extracting the spectra. Positions from
near-IR in 48 out of the 49 objects with 2MASS coordinates have been
found to agree with the X-ray nuclear position within the X-ray radii
of the sources, consistent with the astrometry accuracy provided by
these data. The offset for NGC\,4696 is 7.5'', about the double of the 
size of the nuclear extraction, but NGC\,4696 presents a complex
morphology with a number of knots embedded in strong diffuse emission
in the hard and soft X-ray energies, making the identification of the
nuclear source ambiguous.  X-ray radii and offsets are included in
Col. 4 and 5 in Table
\ref{tab:obsdata}.
The spatial regions sampled by the nuclear extractions generally 
cover the innermost 500 pc, but in a few cases (7) they are larger 
than 1 kpc (see column 5 in Table \ref{tab:catdata}); nevertheless, the 
number of these more distant objects is evenly distributed between AGN 
and SB candidates (see below) and therefore no bias in the results is 
expected due to this effect.

The background region is defined either by a source-free circular annulus 
closed to the nuclei (cases (1) and (3)), or by several circles around 
the sources (case (2)), in order both to take into account the spatial 
variations of the diffuse emission and to minimize effects related to 
the spatial variation of the CCD response. For each source, we extracted 
spectra from each of the datasets. Response and ancillary response 
files were created using the CIAO {\sc mkacisrmf} and {\sc mkwarf} tools.

\begin{figure}
  \caption{NGC\,3379. (left): Smoothed X-Ray image from 0.9 to 1.2 keV and (right): 
the same image with K-band contours from 2MASS over-plotted.}\label{fig:contour}
\end{figure}

The spectra were fitted using XSPEC v.11.3.2. To be able to use
the $\rm{\chi^{2}}$ as the fit statistics, the spectra were binned to
give a minimum of 20 counts per spectral bin before background
subtraction. The task {\sc grppha} included in FTOOLS software has
been used for this purpose. 23 of the 51 objects in our sample fulfill
this criterion (hereafter the Spectral Fitting (SF) subsample). In
the spectral fitting we have excluded any events with energies above
10 keV or below 0.5 keV.

Since our aim is to try to disentangle whether the emission mechanism
might be due to an AGN or to star formation, two models have been
used: a single power-law and a single-temperature optically-thin
plasma emission (MEKAL or Raymond Smith (Raymond \& Smith 1977)
model. For each object five models have been attempted: (1) Power-Law
(PL), (2) Raymond-Smith (RS), (3) MEKAL (ME), (4) PL+RS and (5)
PL+ME. The power-law plus thermal combinations (models 4 and 5) have
been taken into account to include the possibility that the two
emission mechanisms are relevant. We do not expect large differences
between MEKAL and Raymond-Smith models. Furthermore, we have included
a photoelectric absorption law (called 'phabs' in XSPEC software) to
fit the absorbers in the line of sight with a cross section
called 'bcmc' by Balucinska-Church \& McCammon (1992). To decide
which is the best fit model for each nuclear source we have selected
the model that gives $\rm{\chi^{2}}$ reduced closer to 1. The
results from the spectral fittings for the five models are given in Table
\ref{tab:fittings_anex}, where Col. 1 indicates the name; Col. 2 gives
the model Col. 3, 4 and 5 include the column density, spectral index
and temperature; and the $\rm{\chi^{2}}$ over the degrees of freedom
(d.o.f) is indicated in Col. 6. The best model has been chosen as that
with the best $\rm{\chi^{2}}$-reduced statistic. When a combination of
power-law plus thermal model is the best model, we have used {\sc
ftest} task to determine whether the inclusion of an additional
component is needed or a single thermal or power-law model could be a
good fit for the spectrum.  The selected model for
each case has been indicated by an asterisk in Column 2.

\begin{figure}[!h]
\begin{center}
  \caption{The ACIS-S spectrum of NGC\,6240 is shown 
in the top panel. The solid line 
corresponds to a power-law plus Raymond-Smith model. Residuals from the fitting
are presented in the bottom panel.The spectral fitting for all the galaxies in
the SF subsample are shown in the electronic edition.}\label{fig:espectroNGC6240}
 \includegraphics[width=0.65\columnwidth, angle=-90]{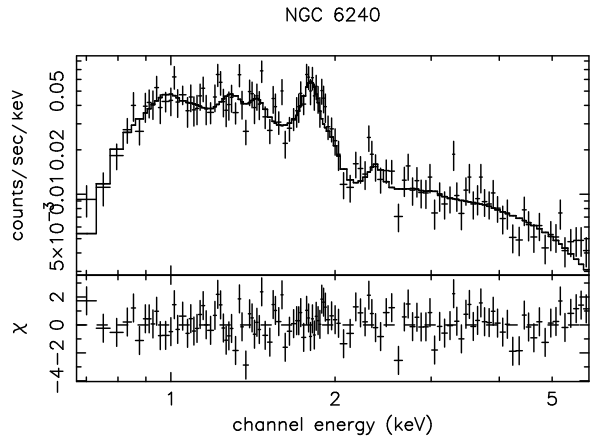}
\end{center}
\end{figure}

NGC\,6240 has been selected as a model example of the process since the
number counts for this source guarantee that the errors in the fitting
due to S/N are minimal. The resulting parameters for the different
fittings of NGC\,6240 are shown in Table
\ref{tab:fittings_anex}. The best fit model is a 
combination PL+RS (Fig. \ref{fig:espectroNGC6240}, top) with the best
$\rm{\chi^{2}}$-reduced statistic (Fig. \ref{fig:espectroNGC6240},
bottom). The soft X-ray spectrum (below 2~keV) shows clear signatures
of thermal emission well described by a optically thin plasma, which
probably originates in a powerful starburst. Strong hard X-ray
emission is also detected and its spectrum above 3~keV is extremely
flat. 

NGC\,4261 has not been included in the SF subsample due to its
spectral complexity that results in unphysical parameters for any of
our five models. The fits are statistically acceptable for all the
remaining objects ($\chi^{2}_{\nu}\sim 1$) except in NGC\,2681
($\chi^{2}_{\nu}=0.54 $) and NGC\,7130 ($\chi^{2}_{\nu}=1.43$). The
resulting fittings are plotted in Fig. \ref{fig:espectroNGC6240} in
the electronic edition for the whole SF subsample.

Only for NGC\,6482 the single thermal model provided a statistically
acceptable fit ($\rm{\chi_{\nu}^{2}\sim1}$). Therefore, in 22 of 23
objects the power-law component is needed to describe the hard
energy spectra observed, indicating a non-negligible non-thermal
contribution in our sample.  Seven objects are described with a
single power-law (NGC\,3690B, NGC\,4374, NGC\,4395, NGC\,4410A,
NGC\,5494, NGC\,4696, NGC\,5746). Although a combination of thermal plus power
law model gives smaller $\chi^{2}$ for NGC\,3690B, NGC\,4410B and NGC\,5746,
the thermal component is not needed to describe their
spectra according to the {\sc ftest}
tool. Furthermore, the same spectral indices are found, including error
bars, assuming a single power-law or a combination
with a thermal model (see Table \ref{tab:fittings_anex}).  Nevertheless,
in 15 out of 22 objects strong residuals remained at low energies
indicating that a single power-law model was not completely
satisfactory. In these cases, a significant improvement was achieved
when a power-law plus optically thin emission was considered,
according to the {\sc ftest} tool.

It has to be noticed that ME+PL and RS+PL with solar metallicity
give the same fitted parameters within the error.

Table \ref{tab:fits} gives the mean (Col. 1), median (Col. 2) and
mean standard deviation (Col. 3) for the logarithm of the 2-10~keV
band luminosity, column density, temperature and spectral index (the
first row for each entry). 
Galactic absorptions can be derived from the HI map (Dickey \& Lockman
1990) using the {\sc nh} tool provided by the HEASARC. The SF sample
of LINERs with enough counts to constrain absorption, showed column
densities exceeding the expected $\rm{N_{H}}$ from
HI map, 
ranging between (0.01-2.87)$\rm{\times 10^{22}~cm^{-2}}$, with a mean 
value of $\rm{3.5 \times 10^{20}~cm^{-2}}$. 
Therefore it is very likely that LINERs generally are much more
absorbed than the Galactic value indicated. The mean 
temperature from the SF subsample is kT=0.64$\pm$0.17~keV while the 
mean spectral index is $\Gamma$=1.89$\pm$0.45. 
The spectral fits provide 2.0-10.0~keV unabsorbed
luminosities for the SF subsample expanding a large range between
$\rm{1.4 \times 10^{38}~erg~s^{-1}}$ and $\rm{1.5 \times
10^{42}~erg~s^{-1}}$ with a mean value of $\rm{1.4 \times
10^{40}~erg~s^{-1}}$.

In order to get a luminosity estimation of the whole sample, we have
obtained a count rate to flux conversion factor between 2.0--10.0 keV,
assuming a power-law model with spectral index of 1.8 and the Galactic
interstellar absorption ($\rm{3 \times 10^{20}~cm^{-2}}$). In
Fig. \ref{fig:lumcomp} the estimated 2.0--10.0~keV luminosity of the
SF subsample is plotted ($\rm{L_{estimated}}$) against the value
obtained from the direct integration of the spectra ($\rm{L{_{fitted}}}$). The luminosities are well
correlated, always less than a factor of 3 within the real
luminosity. Monte Carlo simulations have granted a confidence
of the proposed calibration at 95\% level. We have therefore inferred
a self-consistent estimate of the 2.0--10.0 keV luminosities for the
whole sample, using the SED fitting for the SF subsample and from this
calibration otherwise. In Table \ref{tab:lumflux} we list the 2--10
keV fluxes (Col. 2) and unabsorbed luminosities (Col. 3) of the
nuclear sources for the whole sample, using the empirical calibration
(denoted by `e' in Col. 4) or the spectral fitting (denoted by `f' in
Col. 4). In the cases where the flux and luminosity have been
obtained from the spectral fitting the estimated errors are also
included.

\begin{figure}[!h]
\begin{center}
  \caption{Luminosities estimated  assuming a power-law with an spectral
    index of  1.8 (Log(Lx\_estimated)) versus luminosities computed through the
    spectral fittings (Log(Lx\_fitted)). Objects with the same
    results with both methods should be in the continuous line;
    dashed lines are luminosities from our estimate 3 times higher
    and lower than the luminosities from spectral fitting,
    respectively (Log(Lx\_estimated)=Log(Lx\_fitted)$\rm{\pm}$0.48). SF subsample. }
\label{fig:lumcomp}
\end{center}
\end{figure}


\subsubsection{Image Analysis}

In order to gain insight into the emission mechanisms in the whole
sample we studied the X-ray morphology of the sources in six energy
bands: 0.6--0.9, 0.9--1.2, 1.2--1.6, 1.6--2.0, 2.0--4.5, and
4.5--8.0~keV. The bands were chosen in order to maximize the detection
as well as to obtain a good characterization of the spectra, as it is
illustrated in the next Section. In the last energy band
(4.5--8.0~keV), the range from 6.0 to 7.0~keV has been excluded to
avoid the possible contamination due to the FeK emission line (the
corresponding band will be called (4.5--8.0)$\rm{^{*}}$ hereafter).
The most common emission features in the 2-10~keV band of AGN spectra
are those of iron between 6.4--6.97~keV (depending on the ionization
state of Fe), related with the reflexion in the accretion disk. 
Only 5 objects (NGC\,5194, UGC\,08696,
NGC\,6240, NGC\,7130 and UGC\,05101) have a point like source after
continuum extraction. Although UGC\,08696 shows a compact nuclear
source in this energy band it can not be directly associated with a
FeK line because it has a broad high energy component (See Appendix of
UGC\,08696 for details). 
 
The images were adaptively smoothed (the smoothing depends on the
count rate of the pixel vicinity) with the CIAO task {\sc csmooth},
using a \emph{fast Fourier transform} algorithm and a minimum and
maximum significance signal-to-noise ratio (S/N) level of 3 and 5,
respectively. Smoothing algorithms are useful when the count-rate of
the diffuse emission is closed to the background level. Adaptively
smoothed images were not used for any quantitative analysis, but only
for a morphological classification. The images in the four bands
0.6--0.9, 1.6--2.0, 4.5--8.0$^{*}$ and 6--7 keV are given in
Fig. \ref{fig:clasif} (for all the galaxies see the electronic
edition).

Since we focus our attention in the nuclear sources, no attempt has
been made to fully characterize the flux and the spectral properties
(when possible) of the extra-nuclear sources, whose study is out of the
scope of the paper. As a first insight into the nature of LINERs 
we have taken the existence of an unresolved compact nuclear source in the 
hard band (4.5--8.0~keV) as evidence for an AGN. 
Of course, detection of broad emission lines
at multi-wavelength observations will be needed to asses their nature.
The sample has been grouped into 2 categories:

\begin{itemize}
\item \textbf{AGN candidates:} We include all the galaxies with a
clearly identified unresolved nuclear source in the hard band
(4.5--8.0)* keV. 
In Fig. \ref{fig:clasif}a we show NGC\,4594, as an example of AGN
candidate, where there exists a clear point-like source in the hardest
band (centre-left). 59\% (30/51) of our sample galaxies 
have been classified as AGN-like
nuclei; the median luminosity is
$\rm{L_{X}(2-10~keV)=1.2 \times 10^{40}~erg~ s^{-1}}$, whereas
it is $\rm{L_{X}(2-10~keV)=3.8 \times 10^{39} erg~ s^{-1}}$ for
the whole sample (Fig. \ref{fig:histolum}).

\item \textbf{Starburst candidates:} Here we include all the
objects without a clearly identifiable nuclear source in the hard band.
In Fig. \ref{fig:clasif}b we show the images of CGCG\,162-010 as an
example of these systems. Note that there does not appear to be a
nuclear source in the hardest energy band (centre-left). 41\% (21/51) of
the sample of LINERs falls into this category. The median
luminosity is $\rm{L_{X}(2-10~keV)=1.7 \times 10^{39}~erg~ s^{-1}}$
(Fig. \ref{fig:histolum}).
\end{itemize}

\begin{figure}[!hbp]                   
\begin{center}
\caption{Images (a) for the AGN candidate NGC\,4594 and (b) for the SB candidate
CGCG\,162-010 (b). The top image corresponds to the 0.6-8.0~keV band
without smoothing.  The following 4 images correspond to the X-ray
bands 0.6--0.9 (centre-left), 1.6--2.0 (centre-centre), 4.5--8.0$^{*}$
(centre-right) and 6.0--7.0~keV (bottom-left). The 2MASS image in Ks
band is plotted in the top box. The enlarged view of the region marked
as a rectangle in the centre-right image is the sharp-divided HST
optical image in the filter F814W (bottom-right). All the galaxies are
presented in the electronic edition. }
\label{fig:clasif}
\end{center}
\begin{center}
\end{center}
\begin{center}
(a) NGC\,4594
\end{center}
\begin{center}
\end{center}
\begin{center}
(b) CGCG\,162-010
\end{center}
\end{figure}

\begin{figure}[!h]
\begin{center}
\caption{Luminosity (2-10 keV) histogram for our whole sample (empty histogram),
objects classified as AGN candidates (grey histogram) and for SB candidates
(dashed histogram). Median values are included.}
\label{fig:histolum}
\end{center}
\end{figure}

The classification of each object is included in Col. 6 in Table
\ref{tab:lumflux}. The histogram of the derived X-ray luminosities
for the two groups are presented in Fig. \ref{fig:histolum}. The
median luminosity is higher for AGN like nuclei ($\rm{\sim}$ 10 times)
but there exists a clear overlap in the range
($\rm{[10^{37}-10^{42}]erg~s^{-1}}$). We do not find clearcut
difference in luminosities between AGN and SB candidates.

With respect to the SF subsample (see Table \ref{tab:fits}), the
spectral index and its standard deviation in AGN-like (18/23) is
$\Gamma = \rm{1.7\pm0.3}$ and the temperature is $kT= \rm{0.6\pm0.2}$
keV. In the SF subsample 5 objects have been classified as SB-like
nuclei (namely, NGC\,4438, NGC\,4696, CGCG\,162-010, NGC\,5846 and
NGC\,6482). Excepting NGC\,6482, fitted with a single thermal model,
all the objects have been fitted including a power-law component.  The
origin of this power-law contribution coming from the presence of an
AGN, remains therefore as an open possibility for our sample
objects. A full discussion at this respect is made in Section
\ref{discussion} and in the Appendix for individual sources.



\subsubsection{Color-Color Diagrams}

We have explored the possibility of using X-ray colors to obtain
information about the emission mechanism in these objects. Previous
works have explored this possibility in an AGN sample (Ceballos \& Barcons 1996),
X-ray surveys of galaxies (Hasinger et al. 2001), X-ray source populations 
in galaxies (Grimm et al.
2005; Heinke et al. 2005) or diffuse emission in star-forming galaxies
(Strickland et al. 2004). We have built color-color diagrams ,
using different hardness ratios of the form Q$\rm{_i}$ =
(H$\rm{_i}$-S$\rm{_i}$)/(H$\rm{_i}$+S$\rm{_i}$) (i = A, B, C), with the same bands used
for imaging classification: S$\rm{_A}$=0.6-0.9~keV, H$\rm{_A}$=0.9-1.2~keV ($\rm{Q_A}$),
S$\rm{_B}$=1.2-1.6~keV, H$\rm{_B}$=1.6-2.0~keV ($\rm{Q_B}$), and
S$\rm{_C}$=2.0-4.5~keV, H$\rm{_C}$=(4.5-8.0)$\rm{^{*}}$~keV ($\rm{Q_C}$).
These six bands defining three hardness ratios were selected after
verifying that they are specially well suited to maximize the
differences between a thermal plasma model and emission in the form of
a power-law; while at the low energy range (below 2~keV), the thermal
contribution becomes more significant, at higher energy ranges a
relatively larger flux is expected whenever the power-law contribution
is needed.  The use of the same bands as for the imaging analysis
allows a direct quantification of the properties of the nuclear
sources used for morphological classification.

Counts were extracted directly from the source and background event
files described for the spectral extraction. Error bars were computed
as one standard deviation in the count rates. 
Hardness ratios were calculated for all the bands in which the measured 
error in the count rates were less than 80\%.

\begin{figure}[!h]
  \caption{Color-color diagrams for a RS model (light grey filled
  grid), PL model (dark grey filled grid) and combination model (empty
  grid) for $\rm{N_{H}=10^{20}~cm^{-2}}$ (continuum line) and
  $\rm{N_{H}=10^{22}~cm^{-2}}$ (dashed line). Light-grey triangles
  are AGN-like objects and dark-grey circles are SB-like objects. 
  NGC\,6240 is plotted as a black star. The grids were
  calculated for $\rm{\Gamma=0.4~to~2.6}$ (from up to down) for PL model, for
  $\rm{kT=0.4~to~4.0~keV}$ (from down to up in the thermal model and from 
  left to right in the combination model) for the RS model and
  $\rm{N_{H}=(0.5-30.)\times 10^{20}~cm^{-2}}$ (from left to right) in the single model
  case.}\label{fig:diagrams}
\includegraphics[width=0.49\textwidth]{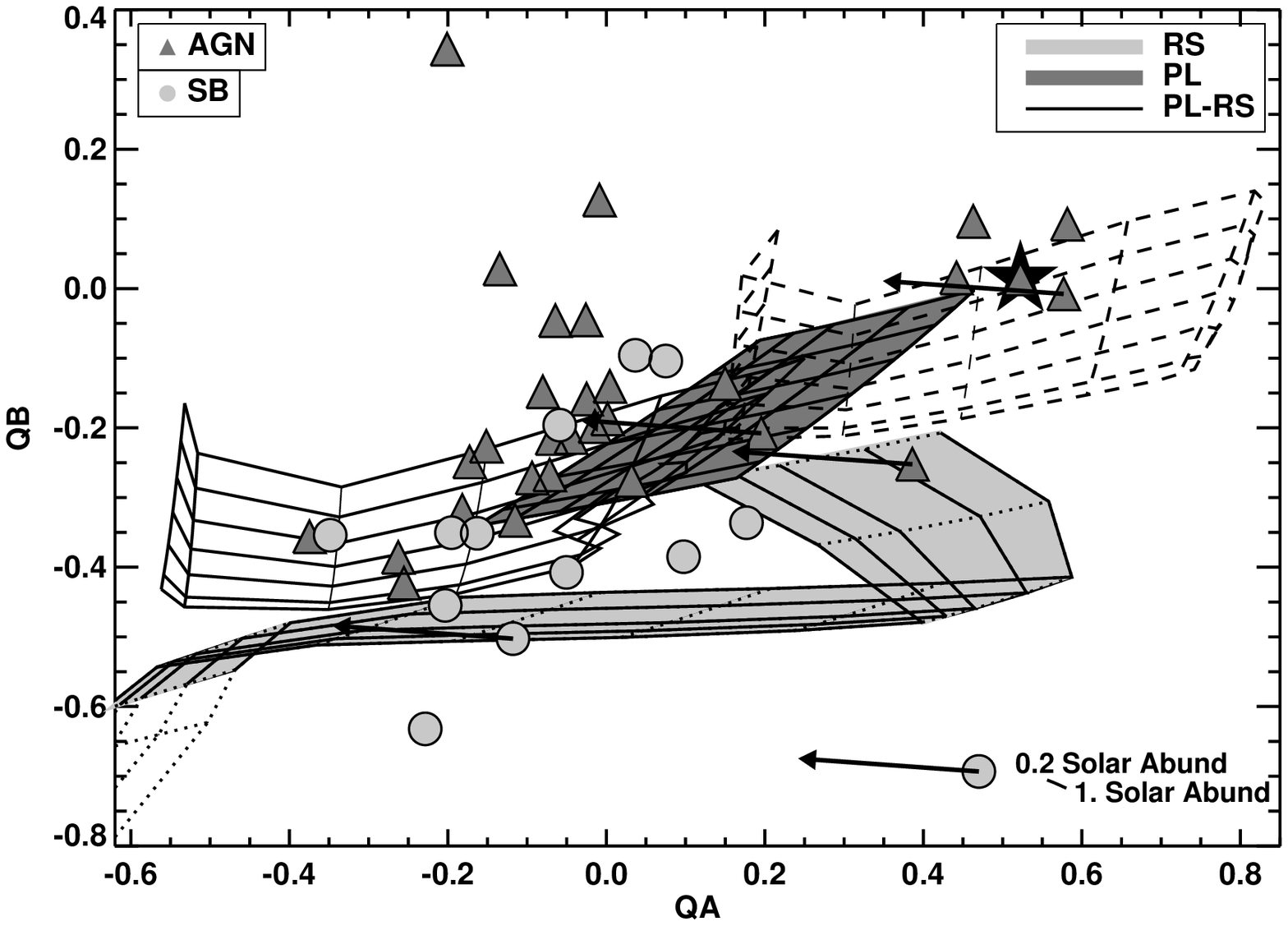}
\includegraphics[width=0.49\textwidth]{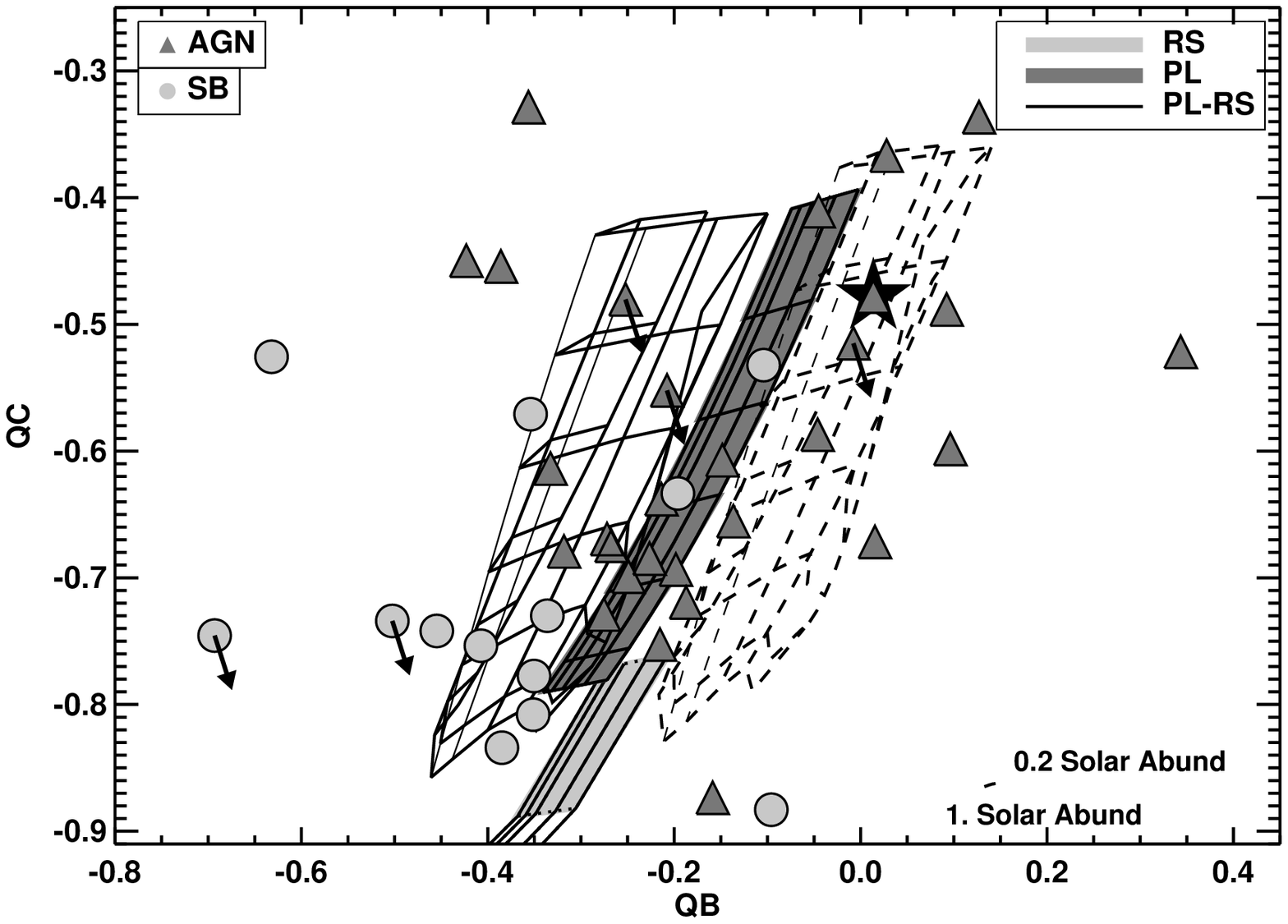}
\includegraphics[width=0.49\textwidth]{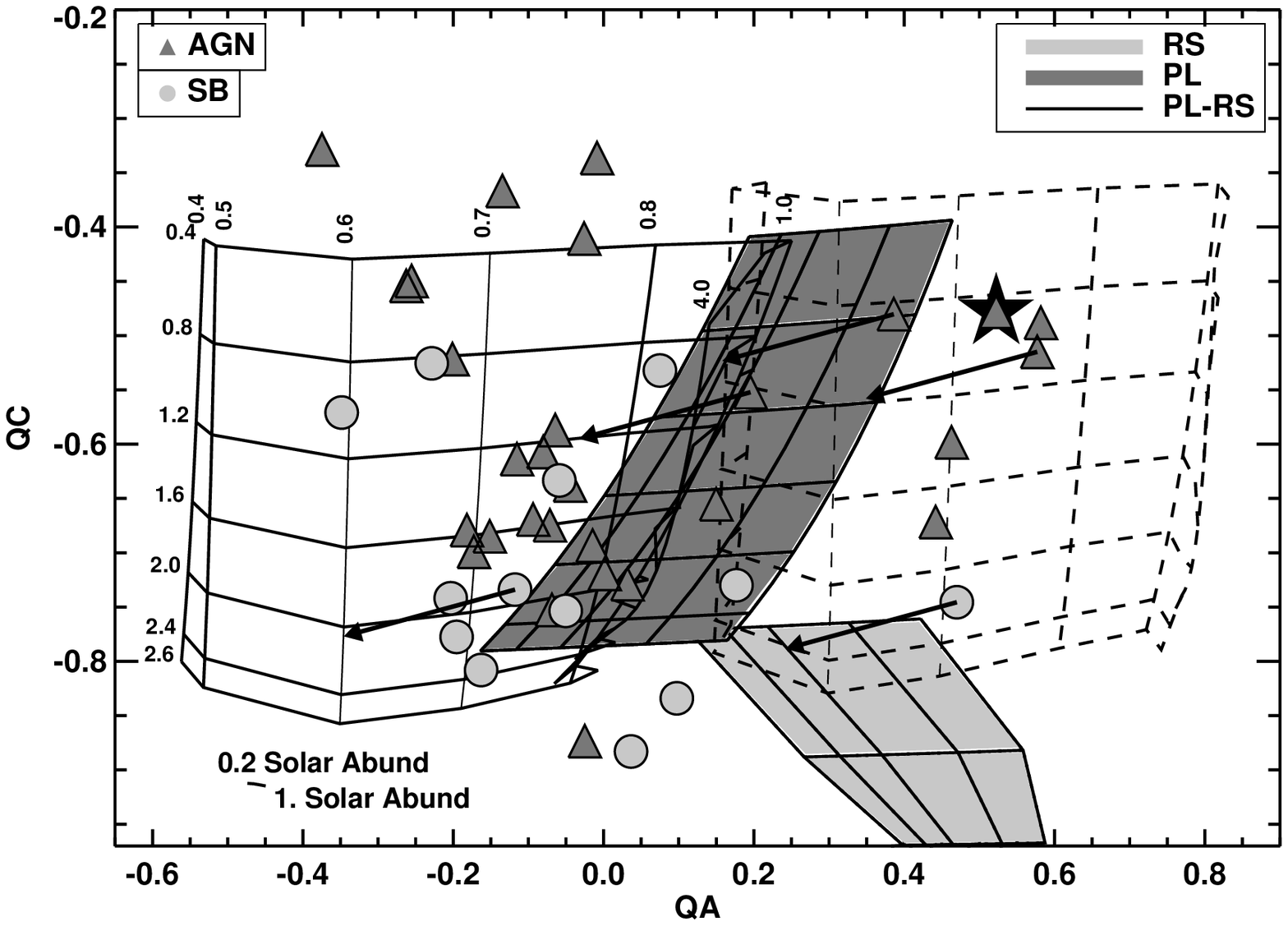}
\end{figure}

In Fig. \ref{fig:diagrams} we plot $\rm{Q_A}$ versus $\rm{Q_B}$
(top), $\rm{Q_B}$ versus $\rm{Q_C}$ (centre), and $\rm{Q_A}$ versus
$\rm{Q_C}$ (bottom). To compare the observed X-ray colors of the
sources with different spectral shapes, we have computed the colors of pure
power-law (dark grey grid), Raymond Smith model (light grey grid) and a
combination of both models (empty grid). 
In both models, photoelectric absorption by cold gas was included. For
each column density ($\rm{N_{H}}$) and model parameter ($\rm{\Gamma}$
or kT) pair, XSPEC generate a model spectrum that was then multiplied
by the effective area at each energy (obtained from the response
matrices for the actual data) and sampled appropriately. The output
was thus a model of the number of photons detected per second as a
function of energy, which could be compared with the
observations. These simulated data were then used to calculate the
hardness ratio of ACIS-S observations as a function of $\rm{N_H,~\Gamma}$ 
and/or kT.  Grids of points were determined for parameters $\rm{\Gamma=0.4-2.6}$ 
and kT~=~0.1--4.0~keV for the power-law (Fig. \ref{fig:diagrams} dark grey
grid) and Raymond Smith (Fig. \ref{fig:diagrams} light grey grid)
models, respectively, and for $\rm{N_H=[1-30]\times 10^{20}~cm^{-2}}$
in the single models. Several grids have been computed in the
combination model with $\rm{N_H=[10^{20}, 10^{21}, 10^{22},
10^{23}]~cm^{-2}}$, with 50\% of contribution from both models at 1~keV. 
Note that variation in the column density in the
combination model result in a shifting of the grid up and right, with
higher values at low energies (Fig. \ref{fig:diagrams} empty grid).
This effect is smaller for low $\rm{N_H}$, but it is quite noticeable 
for higher column densities ($\rm{N_H > 10^{22}}$~cm$\rm{^{-2}}$).  In Fig. 
\ref{fig:diagrams}, two combination models are given as empty grids 
for N$\rm{_H=10^{20}}$~cm$\rm{^{-2}}$ (left) and $\rm{10^{22}}$~cm$\rm{^{-2}}$ (right, 
indicated with dashed lines), respectively.  Therefore, these
grids could help to disentangle whether these objects are strongly
obscured. For less energetic colors, both absorption effects and
thermal contribution become more important, thus both column density
and temperature could be better estimated. The highest energetic color
($Q_{C}$) is better suited to estimate the power-law
contribution. Note that the grid of models have been computed for
ACIS-S observations but for the objects observed with ACIS-I
(namely, NGC\,3608, NGC\,3690B, NGC\,4636, NGC\,5746 and NGC\,6251) an arrow
has been included to represent the correction to be done to take into account 
the different
sensitivities. This correction is important at low energies
(i.e. $\rm{Q_{A}}$) where the different sensitivity of both instruments
become more important. The parameters have been estimated with this
correction.

It has to be noticed that, although our
grid of models has been computed for solar abundances, the 
variation between solar abundance and 0.2 bellow solar abundance is always 
smaller than other effects (see Fig. \ref{fig:diagrams}).

In Fig. \ref{fig:diagrams} we plot $\rm{Q_A}$ versus $\rm{Q_B}$ (top),
$\rm{Q_B}$ versus $\rm{Q_C}$ (centre) and $\rm{Q_A}$ versus $\rm{Q_C}$
(bottom) for the subset of 42 sources for which the three $\rm{Q_{i}}$ 
values are available. The resulting $\rm{Q_A}$, $\rm{Q_B}$ and $\rm{Q_C}$ 
values are given in Table \ref{tab:counts}. SB-like nuclei are plotted
as dark grey circles and AGN-like nuclei as light grey triangles in
these diagrams. 

The reliability of this hardness ratio diagrams can be tested
by comparing their results with those given by the more conventional
method of fitting models to the observed spectra. For this
comparison, we used first the bright source NGC\,6240 (black star in
Fig. \ref{fig:diagrams}), since we have determined the spectral
parameters from the spectral fitting with high reliability. 
Considering the error bars, the spectral index estimated from
color diagrams is $\rm{\Gamma}$ = [0.8-1.0], the temperature is kT =
0.7-0.8~keV and the column density may be high (N$\rm{_H\sim 10^{22}
cm^{-2}}$). From the spectral fitting parameters we have obtained
$\rm{\Gamma=1.03_{-0.15}^{+0.14}}$, kT =
$\rm{0.76_{-0.06}^{+0.06}}$~keV and
$\rm{N_{H}=1.1_{-0.1}^{+0.1}\times10^{22}cm^{-2}}$, well within the
range of values provided by the estimation from the color-color
diagrams.

The method has then been tested with the SF subsample. 
The parameters estimated from color-color diagrams are included in 
Table \ref{tab:fittings}, where the name is shown in Col. 1; SB/AGN
classification is shown in Col. 2; the chosen model is included in
Col. 3 for the objects in the SF subsample, the absorbed column
density from the fitted spectra (Col. 4) or color-color diagram
estimation (Col. 5); as well as power-law index (Cols. 6 and 7); and
temperature (Cols. 8 and 9). Note that the column density estimation
from color-color diagrams has been included as ``$\sim 10^{22}$'' only
those cases where color-color diagrams provide a clear indication of
high obscuration.

\begin{figure}[!h]
  \caption{Temperature (top) and spectral index (bottom) 
  comparison between estimated from color-color diagrams values
  and fitted values. Objects with large departures from the fitted 
  value have been included with the names. Details are shown in 
  the text.}\label{fig:compar-kt-index}
\includegraphics[width=0.49\textwidth]{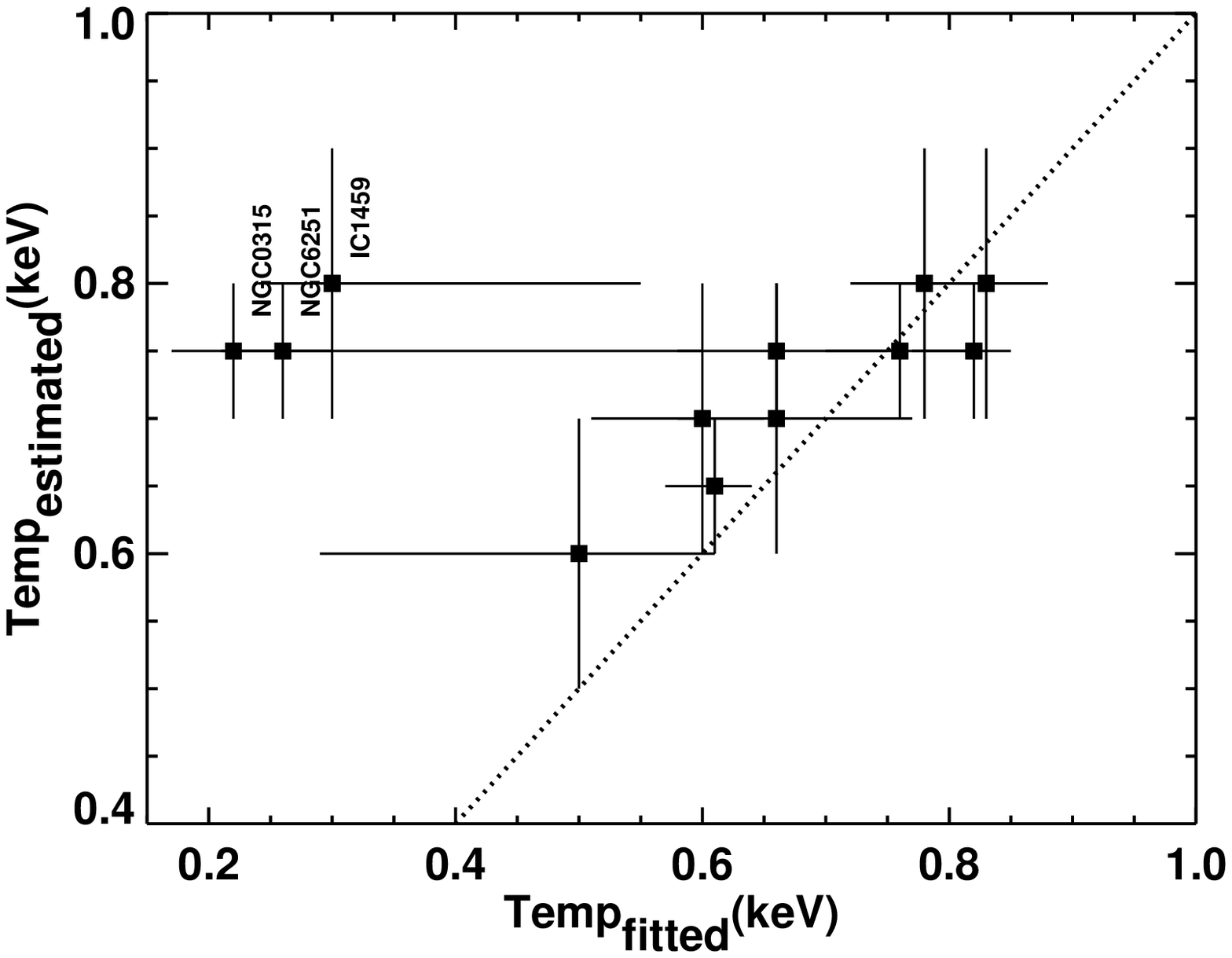}
\includegraphics[width=0.49\textwidth]{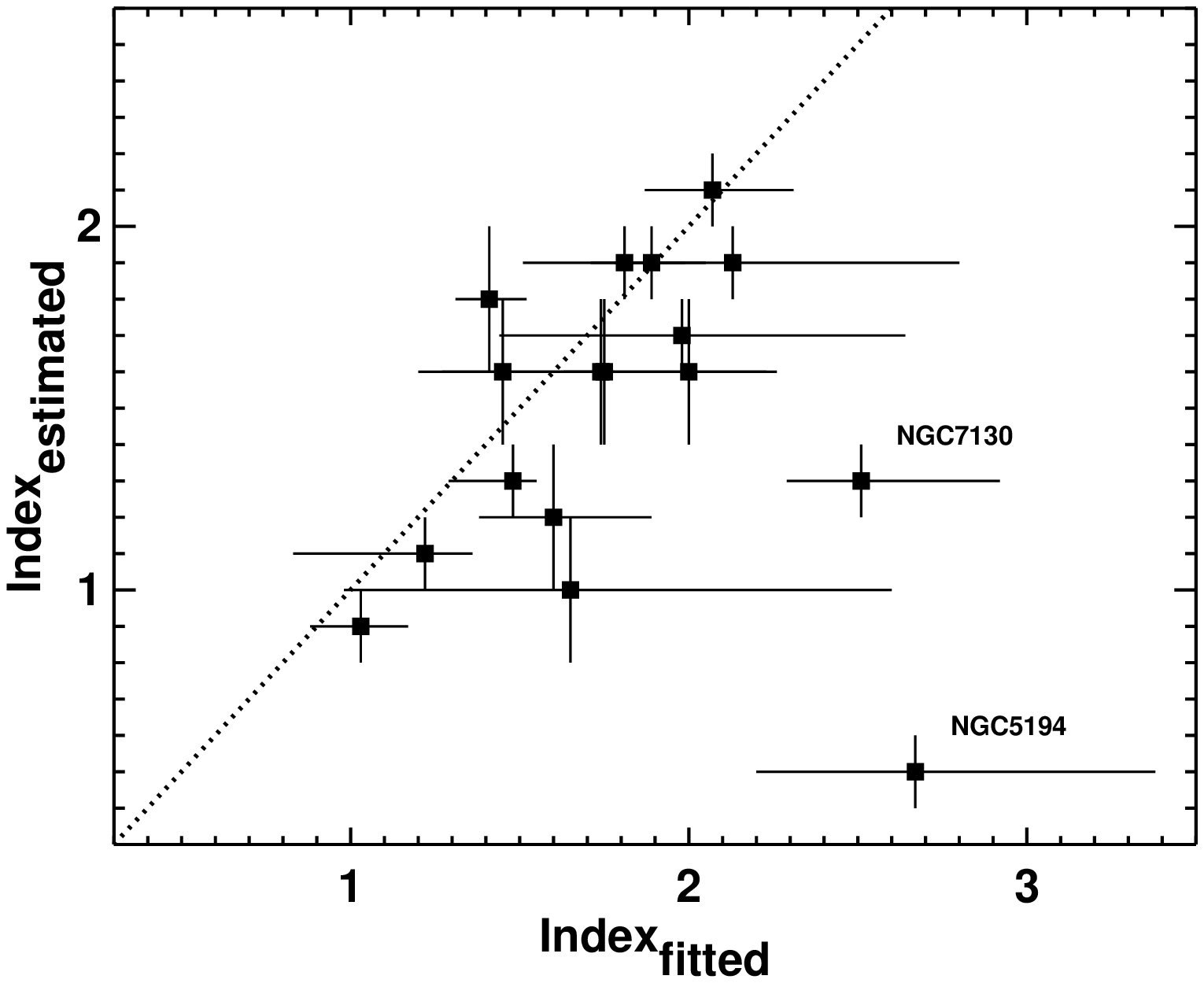}
\end{figure}

Seventeen objects (12 SB-like and 5 AGN candidates) do not have
any estimation from the color-color diagrams. 9 of them have large
errors in the highest energy band, and consequently $\rm{Q_{C}}$ is
not reliable enough. The remaining 8 objects (ARP\,318B, NGC\,1052,
NGC\,4395, NGC\,4494, MRK\,266NE, UGC\,08696, NGC\,5866 and NGC\,6482)
fall out from at least two of the three color-color diagrams:
NGC\,1052, MRK\,0266NE and UGC\,08696 have a too high value for
$\rm{Q_C}$ ($\rm>0.3$), most probably due to the contamination of
jet-related X-ray emission (see Appendix on the individual sources).
ARP\,318B, NGC\,4494 and NGC\,6482 are out of the grid because of low
values of $\rm{Q_C}$; NGC\,6482 is the only galaxy resulting to be
best fitted by a single thermal component, and its position in the
color-color diagrams is indeed closer to the thermal model. The
situation for the other two galaxies, NGC\,4395 and NGC\,5866, is
however less clear.

Fig. \ref{fig:compar-kt-index} (top) shows the comparison of the 
temperature obtained from the spectral fitting with that estimated
from color-color diagrams. Excepting NGC\,0315, NGC\,6251 and IC\,1459
that show temperatures estimated from color-color diagrams much higher
than the fitted ones, it can be concluded that color-color diagrams
provide a good temperature estimation.  NGC\,0315 has been fitted by
Donato et al. (2004) with kT=0.51~keV, much closer to our estimated
temperature. Considering the extended structure and the jet-like
emission in this galaxy, differences in the background subtraction methods may
explain such a discrepancy.
Note that NGC\,6251 and IC\,1459 are the only galaxies for which
PL$+$RS and PL$+$MEKAL fitting provide very different values of kT;
even if PL$+$MEKAL model is selected in these cases, the statistical
estimator is not so sharp and, if a PL$+$RS is considered, the
resulting temperatures are in good agreement with the estimated ones.

Fig. \ref{fig:compar-kt-index} (bottom) shows the spectral index
correlation between the values from the spectral fitting versus those
estimated from color-color diagrams. The spectral indices estimated from
the color-color diagrams result to be somewhat underestimated, but
within less than 40\% of the fitted values, excepting for 
NGC\,5194 and NGC\,7130, for which the fitted values are rather unrealistic.

All the estimated column densities corresponds with the fitted
ones.

Therefore, it seems reasonable to conclude that the information provided 
by color-color diagrams can be used as a rough estimate of the physical 
parameters describing the X-ray SED of our targets, this method being 
specially valuable for the objects where no spectral fitting can be made.

For the whole sample, the main general conclusion from the diagrams in
Fig. \ref{fig:diagrams} is that the spectra of LINERs cannot be
described in general by a single power-law. At least two components
are needed: a hard power-law and a soft component represented by a
thermal model. We find that the best fit MEKAL model for a soft
component in most of LINER galaxies has kT between 0.6-0.8~keV 
and it appears quite clearly that
an AGN power-law component may be needed for most of them, since 
$\rm{Q_C}$ values are in general too high for a thermal emission. In
particular, AGN-like nuclei show systematically  high $\rm{Q_C}$
values; SB-like nuclei mostly fall in the power-law region as well. 
This result further emphasizes
that the AGN contribution in our sample of LINER galaxies might be
rather important. 





\subsection{HST-imaging analysis}

To gain some insight into the nature of the X-ray emission, especially
for the LINERs without a detected compact component, we have searched
for the optical counterpart of the X-ray compact sources.  The high
spatial resolution provided by HST observations is needed for our
purposes. In order not to make our analysis dependent on the model
used for describing the underlying host galaxy, we have decided to
remove such a contribution by applying the sharp-dividing method to
the original HST images. This technique has been proved to be
very well suited to remove large-scale galactic components and
therefore is a very convenient way to look for subtle, small-scale
variations and discuss the possible presence of both dust extinguished
and more luminous regions (Sofue et al. 1994; M\'arquez \& Moles 1996;
M\'arquez et al. 2003 and references therein; Erwin \& Sparke 1999;
Laine et al. 1999).  It consists on applying a filtering to the
original image, with a box size several times that of the PSF FWHM, and then
dividing the original image by the filtered one.
The resulting so-called sharp-divided images are
plotted in Fig.\ref{fig:clasif} (bottom-right).  We note that absolute
astrometry has been performed by using the position of the coordinates
of those objects in the images cataloged by USNO, excepting NGC\,4410,
NGC\,4596 and NGC\,4696 for with the position of the central galaxy
had to be taken from the RC3 catalogue due to the lack of reference
stars.

According to the appearance of the central regions, two main groups
can be distinguished: (a) galaxies with compact nuclear sources, 
when a knot coincident with the central X-ray source is detected in
the sharp-divided image (see for instance Fig. \ref{fig:clasif} for
NGC4552) (C in Column 7, Table \ref{tab:lumflux}); 35 objects fall
into this category; and (b) galaxies with dusty nuclear regions, 
when dust-lanes like features are seen but no central knot is detected
(see for instance Fig. \ref{fig:clasif} for NGC4438) (D in Column 7,
in Table \ref{tab:lumflux}) with 8 objects belonging to this
class. Two galaxies, namely NGC\,3608 and NGC\,4636, cannot be
classified because of saturation or rather low S/N.

\section{Discussion}\label{discussion}

According to their X-ray morphologies in the hard band
(4.5-8$^{*}$~keV), the LINER galaxies in our sample have been
classified into two broad main categories.  We have defined AGN-like
nuclei as those objects displaying a hard nuclear point source, 
coincident with 2MASS position of the nucleus. Of the 51 galaxies in
the LINER sample, 30 meet this criterion (59\%). We called SB
candidates those lacking a hard nuclear source because either 1) they lack
an energetically significant AGN or 2) contain highly obscured AGN
with column densities larger than $\rm{\sim 10^{23} cm^{-2}}$.  Ho et
al. (2001) classified the X ray morphology into four categories:
class (I) objects exhibit a dominant hard nuclear point source
(2-10~keV); class (II) objects exhibit multiple hard off-nuclear point
sources of comparable brightness to the nuclear source; class (III)
objects reveal a hard nuclear point source embedded in soft diffuse
emission; and class (IV) objects display no nuclear source.  Dudik et
al. (2005) analyzed a sample of IR-Bright LINERs and classified the
objects following the same scheme.  Considering that we can assimilate
classes (I) and (III) into our group of AGN-candidates and that their
class (II) is equivalent to our SB candidates, they obtained similar
percentages of AGN-like objects since 51\% (28/55) of their LINERs
fall into classes (I) and (III), and 13\% (7/55) were classified as
belonging to class (IV). A detailed comparison for the 18
galaxies in common show that 15 objects share the same
classification. The odd classification for the remaining three nuclei
(NGC\,3628, NGC\,4696 and CGCG\,162-010) can be attributed to our use
of a harder X-ray band for the detection of unresolved nuclear
sources.

As seen in Fig. \ref{fig:histolum}, the 2-10~keV luminosities of the
AGN-like in our 
LINER sample range from $\rm{10^{37}erg~s^{-1}}$ to 
$\rm{10^{43}erg~s^{-1}}$, while
SB-like luminosities range from $\rm{10^{37}erg~s^{-1}}$ to
$\rm{10^{42}erg~s^{-1}}$, both classifications overlapping in the
range $\rm{10^{37}erg~s^{-1}}$ to $\rm{10^{42}erg~s^{-1}}$. The
2-10~keV X-ray luminosities were calculated fitting the SED between
0.5-10~keV in the SF subsample, and assuming a generic power-law model
with photon index $\rm{\Gamma=1.8}$ and the Galactic absorption
otherwise. A similar method was proposed by Ho et al. (2001) who use
an empirical estimate from the count rates.  Satyapal et al. (2005),
by using Ho estimate of the luminosity, have found for the AGN LINERs
a range in luminosities from $\rm{\sim 2
\times 10^{38}}$ to $\rm{\sim 1\times 10^{44}erg~s^{-1}}$.  For the 31
objects in common, the resulting luminosities agree within the errors
but for 3 objects. The differences can be attributed to the presence
of bright FeK lines in the spectra of NGC\,5194 and NGC\,6240; this
is clearly the case for NGC\,6240, with the highest luminosity they
derive (Vignati et al. 1999 obtain $\rm{\sim 1\times
10^{44}erg~s^{-1}}$ once the FeK line contribution is included in the
fit as a gaussian component). For NGC\,3245 the luminosity we estimate
is very much fainter that the one given by them; we however stress
that our estimation is in very good agreement with that by Filho et
al. (2004) (see Appendix).

We have found that LINER X-ray SED can be interpreted as a combination
of a soft thermal component with temperatures of about 0.6-0.8 keV,
maybe due to circunnuclear star formation, and a hard power-law
produced by an obscured AGN.  Regarding the soft component, Ptak et
al. (1999) based on ASCA data found that similar models apply to
starbursts usually showing a temperature greater than 0.6~keV. 
This result is confirmed with \emph{Chandra} observations for starburst
galaxies by Ott et al. (2005) and Grimes et al. (2005), while Teng et
al. (2005) found that the best fit MEKAL model for the soft component
in Seyfert 1 galaxies has kT$\rm{\sim}$~0.25~keV. They discussed that the
low temperatures of the Seyfert 1 galaxies suggest that starburst
activity may not be the dominant energy source of the soft
component. The same conclusion was drawn by Boller et al. (2002) for
F01572+0009. LINER nuclei have similar values than we expect in
starbursts, indicating that star formation activity may be the
dominant energy source of the soft component.  Except in 2
objects (NGC\,0315 and IC\,1459), the temperature obtained from the
spectral fit is about 0.6~keV, therefore hinting that a starburst
component can be inferred in most cases. The spectral index in the
sample galaxies for which the power-law component is needed to fit
spectra (22 objects), range from $\rm{\sim}$1.0 till 4.3 with a mean value
of 1.9 (see Table \ref{tab:fits} for detailed analysis).  In 3 out of
22 objects (NGC\,4494, NGC\,5746 and NGC\,6240) the slope tend to be
significantly flatter ($\rm{\Gamma <1.4}$) than what is typically
observed in AGNs. Ptak et al. (2003) explain this flattening in a
Compton thick scenario as due to a combination of reflection from
optically thick central material, scattering from optically thin
(unlikely highly ionized) material and leakage of X-ray through
patches in the obscuring material. 
Nevertheless, we
recall that simple models have been used to test the need of power-law
and/or thermal components to fit these spectra and some artifacts may
be consequently produced for more complex situations where the
presence of several components is absolutely needed to explain the
spectral features.

Our sample can be compared to X-ray observations of nearby
Seyfert 2 galaxies.  In the recent survey of nearby Seyfert galaxies
taken from the compilation by Ho et al. (1997), Cappi et al. (2006)
analyze \emph{Chandra} data for 27 Seyfert 1 and 2 galaxies (mainly Seyfert
2), obtaining values for the power-law index of 1.56 for type 1
Seyfert and 1.61 for type 2, which are not too far from our quoted
median value for LINERs (1.75 $\rm{\pm}$ 0.45). Their most remarkable
result is that the range of column densities they find for Seyferts is
quite similar to what we find for LINERs, with N$\rm{_H}$ 
between 10$\rm{^{20}}$ and 10$\rm{^{22}~cm^{-2}}$.  Finally, when a thermal
component is needed (in 15 out of the 27 galaxies), kT is found in the
range between 0.2-0.8 keV, as it is the case for our LINERs, although
their spectral fitting technique is quite different from ours: they
fit the spectra between 2-10 keV by a single power-law and then a
thermal component is added to the cases where the extrapolation from the 
power-law to lower energies (below 2 keV) results in a soft excess. The mean
value for their (2-10 keV) luminosities (10$\rm{^{39.8}}$ ergs s$\rm{^{-1}}$ for
Seyfert 2 and 10$\rm{^{41}}$ ergs s$\rm{^{-1}}$ for Seyfert 1) is also in good
agreement with our estimated mean value (10$\rm{^{40.16}}$ erg~s$\rm{^{-1}}$).
These results are at variance with those obtained by Guainazzi et al. (2005),
whith luminosities larger than 10$\rm{^{41}}$ ergs
s$\rm{^{-1}}$ for a sample of 49 Seyfert 2. Note however that their
galaxies are more heavily obscured than those in the analysis by Cappi
et al. On the other hand, the spectral indices are fully consitent with those
reported by Mateos et al. (2005) and Gallo et al. (2006) for galaxies
at larger redshifts. Note that, even if their galaxies are more
luminous more obscured objects, they obtain $\rm{\Gamma}$ values varying
between 1.7 for AGN type 1 and 1.9 for type 2 AGNs.  Hence the
conclusion from this comparison is that it is not clear whether LINERs
in the X-ray represent a lower scaled version of AGN activity. For a
more definitive conclusion a full, homogeneous analysis using the same
methodology and similar selection criteria is therefore needed. At
this respect, we have collected archival X-ray data for a sample of
Seyfert galaxies that will be studied with the same methods we have
used for LINERs, and their analysis will be the subject of a
forthcoming paper.


To gain some insight into the nature of the X-ray emission, especially
for the LINERs without a detected compact component, we have searched
for the optical counterpart of the X-ray compact sources. We note that
all the galaxies classified as AGN by their X-ray imaging show compact
nuclei (28 galaxies). Among those showing SB X-ray morphology (17
objects) 8 galaxies (NGC\,3507, NGC\,3607, NGC\,4438, NGC\,4676A,
NGC\,4676B, NGC\,4698, CGCG\,162-010 and NGC\,5846) show dusty nuclear
sources; 5 out of 8 showing low column densities and 3
(NGC\,3507, NGC\,3607, NGC\,4676B) without estimated column
density. The remaining 7 galaxies classified as SB candidates host
nevertheless compact nuclear sources in the optical counterpart (namely 
NGC\,3245, NGC\,3379, NGC\,4314, NGC\,4459, NGC\,4596, NGC\,4696 and
NGC\,7331). NGC\,3608 and NGC\,4636 remain unclassified. We stress
that NGC\,4696 is obscured at X-rays (as some of the
object in Chiaberge et al. 2005) so in this case the AGN would be
only visible in the optical, i.e., a low optical extinction has to be
coupled with a relatively high X-ray column density; this kind of
situation might be explained by, e.g., unusual dust-to-gas ratios
(Granato et al. 1997) or by the properties of the dust grains
(Maiolino 2001).  Moreover in NGC 4696 a broad balmer H$\rm{\alpha}$ line 
was detected by Ho et al. (1997). NGC\,3245 shows also evidences of 
its AGN nature from the optical, with double-peaked H$\rm{\alpha}$ and [NII]
profiles indicative of the presence of spatially unresolved rapid
rotation near the nucleus (Barth et al. 2001) and the radio
frequencies, where an unresolved nuclear radio source is detected
(Wrobel \& Heeschen 1991). Unresolved nuclear radio cores have been
also found for NGC\,7331 (Cowan et al. 1994) and NGC\,4314 (Chiaberge
et al 2005). For NGC\,4596 Sarzi et al. (2001) has determined a Black
Hole Mass of 7.8 x 10$\rm{^7}$ M$\rm{\odot}$ from an STIS kinematic analysis of
the source. 
Thus it appears that even in the cases we have
called SB candidates, the existence of a mini or an obscured AGN
cannot be completely ruled out, since the region mapped by these
observations may correspond to distances to the centre less closer to
the AGN, who may be much more obscured; they would map regions much
further out instead, what would explain that thermal mechanisms most
probably due to star-forming processes seem to be the dominant energy
source in these systems (Rinn et al. 2005).



However the unambiguous determination of the presence of an AGN
needs of a detailed study of the individual sources, to discard the
eventual mechanisms producing unresolved hard X-ray nuclear
morphologies originated in star forming processes. Eracleous et al. (2002)
argues favoring the SB nature of NGC\,4736 (see Apendix). In particular the
contribution of high mass X-ray binaries (HMXBs) and ultra-luminous
X-ray sources (ULXs)\footnote{See Fabbiano 2005 for a review on the
different X-ray source populations in galaxies.} needs to be
estimated,
since they are expected to dominate the emission of star
forming galaxies (see for instance the data for the Antennae by Zezas
\& Fabbiano 2002). The high X-ray luminosity found for an ULX in the
star forming galaxy NGC 7714 (Soria and Motch 2004), amounting to 6
$\rm{\times}$ 10$\rm{^{40}}$ erg s$\rm{^{-1}}$, evidences how important the
contamination produced by such objects may be if they are found at
nuclear positions, making this analysis extremely difficult and
implying that only indirect evidences can be invoked. Hence we should
keep in mind that contribution from ULXs cannot be discarded out.

Stellar population synthesis will help to investigate the relevance
of the contribution from HMXBs since they are short lived
(10$\rm{^{6-7}}$ years). Data on population synthesis exist for 14 out of
the 51 galaxies by Cid Fernandes et al. (2004) and Gonz\'alez-Delgado
et al. (2004). Only in one galaxy, NGC 3507, a large contribution of
a young population of an age $<$ 10$\rm{^8}$ years ($\rm{\sim}$ 27\%) does exist. For
the remain galaxies the contribution is always lower than 3\%.  Then
it is not clear that HMXBs may be an important ingredient on the nuclear
X-ray emission. It has to be noticed that the mapped region of the
optical data (1.5'') is quite the same as that used in our X-ray analysis.

Data at radio frequencies are also crucial to understand the AGN
character of LINERs. An unresolved nuclear radio core and flat
continuum spectra have been taken as the best evidence for their AGN
nature (see Nagar et al. 2005, Filho et al. 2004 for a full
discussion). 33 out of the 51 objects have been observed at
radio-frequencies (Filho et al. 2000, 2002, 2004, Nagar et al. 2000,
2002, 2005, Falcke et al. 2000). From these 33 objects 16 galaxies
seems to be detected with good confidence, but only upper limits can
be given for 17 objects. 3 out of the 16 radio detected (NGC\,3628,
NGC\,4636 and NGC\,5846) show radio steep spectrum which can be taken
as evidence of their Starburst nature. The remaining 13 galaxies show
AGN nature according to radio diagnosis; 9 of them show X-ray AGN
morphology and 4 have an SB classification (NGC\,3245, NGC\,4459,
NGC\,5866 and NGC\,7331). In principle, compact radio cores should
appear as compact sources at X-ray frequencies unless the X-ray
nucleus is obscured due to large amounts of dust; nevertheless, it does 
not seem to be the case for these 4 SB galaxies with compact radio cores 
since they do not appear to have large column densities. Thus the reason for
the discrepancy needs to be found.  For NGC\,3245 and NGC\,4459 the
AGN radio classification was made by Filho et al. (2004) on
compactness arguments based on 5 arcsecons resolution radio imaging, so
much better resolution data are needed. NGC\,7331 needs also better
data due to the marginal nuclear source identification.  The only
clear radio-AGN is NGC\,5866. A closer inspection of this source (see
our Figure 5) shows that a weak nuclear source can be identified in
the hard band. Then we should conclude based on the available data
that AGN radio cores appear as AGN-like systems at X-ray frequencies.

>From the precedent analysis it is clear that a multi-frequency approach
is needed to get insight into the AGN nature of these galaxies, but
since most of the hard emission between 4.5-8 keV is coming from an
unresolved nuclear source, the natural explanation seems to be that
the X-ray unresolved nuclear source is due to a low luminosity
AGN. Low luminosity AGNs in the nucleus of early type galaxies being
quiescent at optical or UV frequencies have been invoked by Soria et
al (2006a and b) due to the existence of Supermasive Black Holes
(SMBH) in the nucleus of these galaxies and in their X-ray properties.
Therefore since for most of the LINERs a SMBH has been detected
(Satyapal et al. 2005) the AGN nature of the X-ray emission seems to be
appropriate.

\section{Conclusions}

The primary goal of our study was to determine the X-ray nuclear 
characteristics of a statistically significant sample of LINERs which
were selected based on their optical emission properties (Veilleux and
Osterbrock 1987). Archival \emph{Chandra} ACIS data have been used
to study the nature of our sample, with the aim of analyzing whether
the observed X-ray emission is consistent with AGN
powered and whether star forming emission can be ruled out in our
sample or not. We have first classified the nuclear
morphology, according to the compactness in the hard band, as AGN
candidates whenever a clearly identified unresolved nuclear source is
found in the 4.5 to 8.0~keV band and as SB candidates otherwise. 
Color-color diagrams for the whole sample and the spectral analysis for
the SF subsample lead to the following conclusions:

\begin{enumerate}
\item Morphologically, 30 out of 51 of LINERs have been classified as
AGN-like candidates. 
\item Thermal plus power-law model with a median temperature of
$\rm{\sim}$0.6~keV (0.6-0.9~keV from color-color diagrams) and median
spectral index of $\rm{\sim}$1.9 (1.4-1.7 from color-color diagrams),
better reproduces the X-ray energy distribution. This indicates a
non-negligible contribution of a non-thermal component in our sample.
\item Luminosities in the energy band 2-10~keV 
span a large range
between $\rm{1.4\times 10^{38}erg~s^{-1}}$ and $\rm{1.5\times
10^{42}erg~s^{-1}}$ with a mean value of $\rm{1.4\times
10^{40}erg~s^{-1}}$. For the two groups separately, AGN candidates range between 
$\rm{7.8 \times 10^{37}~erg~ s^{-1}}$ and 
$\rm{1.5 \times 10^{42}~erg~ s^{-1}}$ while SB candidates range
between $\rm{8.7 \times 10^{37}~erg~ s^{-1}}$ and 
$\rm{5.3 \times 10^{41}~erg~ s^{-1}}$. The broad overlap
in X-ray luminosities indicates that this parameter can not be used
to discriminate between AGN and SB candidates in LINERs. 
\item We stress that color-color diagrams are a valid tool to roughly
estimate the values of the physical parameters used in the models for
the spectral fitting, providing important clues on the amount of
nuclear absorption. Their use is especially interesting when the
spectral fitting is not possible.
\item The estimated column densities from either the spectral analysis or
color-color diagrams show that it is very likely that LINERs are
generally significantly obscured, especially in SB candidates. This results
is consistent with the hypothesis that they may host an obscured AGN.
\item All the galaxies classified as AGN by the
X-ray imaging show compact nuclei at the resolution of HST images.
8 out of 14 galaxies showing SB
X-ray morphology have dusty nuclear sources, indicating that the
objects can be dust obscured, in agreement with the X-ray analysis.
\end{enumerate}

These new data offer some insight on the long standing dispute about
the nature of the ionizing source in LINER. Although contributions 
from HMXBs and ULXs can not be ruled out for some galaxies, we can 
conclude that for a high percentage of LINERs, 59\% at most, an AGN 
does exist in their nuclei. For the remaining 41\%, their existence cannot be
discarded based on the available data. Diagnostics at other
wavelengths have to be explored in order to determine the possible
contribution of an embedded AGN. Future papers will 
be devoted to study whether a link exists in objects with higher
activity levels by comparing the properties of our AGN-like LINER
nuclei with those of a sample of Seyfert nuclei.




\begin{acknowledgements}

This work has been financed by DGICyT grants AYA2003-00128 and the Junta de
Andaluc\'{\i}a TIC114. OGM acknowledges the financial support by the
Ministerio de Educaci\'on y Ciencia through the spanish grant FPI
BES-2004-5044. MAG is supported by the Spanish National program
Ram\'on y Cajal.  We thank X. Barcons and E. Jimenez-Bailon for
helpful comments, D.W. Kim for useful advices in data reduction 
and J. Cabrera for providing us with the Monte Carlo
simulations. This work benefited from fruitful previous work with
A.C. Fabian and J.S. Sanders.  This research has made use of the
NASA/IPAC extragalactic database (NED), which is operated by the Jet
Propulsion Laboratory under contract with the National Aeronautics and
Space Administration. We acknowledge an anonymous referee for
her/his comments and advices on our work which resulted in a great
improvement of the manuscript.
\end{acknowledgements}

\section{Appendix}

{\bf NGC\,315 (UGC\,597, B2\,0055+30).} The high spatial resolution
provided by \emph{Chandra} imaging allowed the detection of X-ray jets, the
most striking one being that along $\sim$10'' to the NW
(see Fig. \ref{fig:clasif} and Donato et al. 2004, Worral et al. 2003).  Worral et
al. (2003) made use of a 4.67~ks duration ACIS-S \emph{Chandra} image to
report an X-ray luminosity, for our assumed H$\rm{_0}$=75
km~s$^{-1}$~Mpc$^{-1}$, of 5.9 $\times$ 10$^{41}$erg~s$^{-1}$ (2-10~keV) 
and a power-law energy index $\Gamma$=1.4 
seen through a moderate intrinsic column density of N$\rm{_H}$ from
2.3 to 8.2 $\times$ 10$^{21}$cm$^{-2}$ for the nuclear component whose
spectrum is fitted with a single-component absorbed power-law. On the
contrary, Donato et al. (2004), using the same spatial area for the
extraction of the nuclear component (1.98$\pm$0.20''), conclude that
to model the corresponding spectrum two components are required: a
power-law ($\Gamma$=1.56) and a thermal one (apec in XSPEC) with solar
metallicity (kT=0.51~keV, N$\rm{_H}$=0.73 $\times$
10$^{22}$cm$^{-2}$), therefore in very good agreement with our results
by using RS+PL as the best model (see Table
\ref{tab:fittings}). Satyapal et al. (2005) class NGC\,315 as an
AGN-LINER (those displaying a hard nuclear point source, with a
2-10~keV luminosity $>$ 2$\times$10$^{38}$ erg~s$^{-1}$, coincident
with the VLA or 2MASS nucleus); their spectral fitting results in
kT=0.54~keV, N$\rm{_H}$=0.8 $\times$ 10$^{22}$cm$^{-2}$ and
$\Gamma$=1.60, that also agree with ours.
\smallskip

{\bf Arp\,318A and B (Hickson Compact Group 16 A and B).}  No point
sources are detected in neither hard band images (4.5-8$^*$~keV or 6-7~keV), 
but X-ray soft emission is extended (see Fig. \ref{fig:clasif}). Turner et
al. (2001) analyze the 40 ksec EPIC \emph{XMM-Newton} first-light
observations and confirm the presence of an AGN in both galaxies A and
B. Three components were fitted to the EPIC\,X-ray spectrum of {\bf
NGC\,833} (Arp 318B): (1) a power-law for the obscured AGN, with
$\Gamma$=1.8 
and N$\rm{_H}$=2.4 
$\times$
10$^{23}$cm$^{-2}$, (2) an unabsorbed power-law for the radiation
scattered into our line of sight by thin, hot plasma directly
illuminated by the AGN, and (3) an optically-thin thermal plasma with
kT=0.47~keV; the luminosity of the AGN component of 6.2
$\times$ 10$^{41}$ erg~s$^{-1}$ results to be 100 times brighter than
the thermal X-ray emission. The core of {\bf NGC\,835} (Arp 318A)
showed a very similar spectrum, with absorbed and scattered power-laws
indicating a heavily obscured AGN (N$\rm{_H}$=4.6
$\times$
10$^{23}$cm$^{-2}$ and $\Gamma$=2.25
) of 5.3 $\times$ 10$^{41}$erg$^{-1}$ (0.5-10~keV) and a soft thermal
component with kT=0.51~keV contributing at 2\% of the total luminosity.
Total counts for source B being insufficient for both the spectral
analysis and the use of color-color diagrams, we only give the
morphological classification as SB candidates for the two nuclei, in
contrasts with the reported \emph{XMM-Newton} spectral results and an estimation of
$\Gamma$ and kT from color-color diagrams for source A, consistent
with \emph{XMM-Newton} results, but with no indication of high obscuration in our
analysis.
\smallskip

{\bf NGC\,1052.}  The X-ray morphology clearly indicates the presence
of an unresolved nuclear source in the hard bands
(Fig. \ref{fig:clasif}), in agreement with the classification by
Satyapal et al. (2004). Evidence for the AGN nature of this object
was already given with the detection of broad lines in
spectropolarimetric measurements by Barth et al. (1999). Guainazzi et
al. (2000) confirmed that its X-ray spectrum may therefore resembles
that of Seyfert galaxies with the analysis of its \emph{BeppoSAX}
spectrum (0.1-100~keV), for which they derive a very good fit with a
two component model, constituted by an absorbed (N$\rm{_H}$=2.0
$\times$ 10$^{23}$cm$^{-2}$) and rather flat ($\Gamma$
$\approx$1.4) power-law plus a ``soft excess'' below 2~keV. The
corresponding flux in the 2-10~keV is 4.0 $\times$
10$^{-12}$ erg~cm$^{-2}$~s$^{-1}$.  The presence of various jet-related
X-ray emitting regions from a short (2.3~ks) \emph{Chandra} observation,
together with a bright compact core and unresolved knots in the jet
structure as well as an extended emitting region inside the galaxy
well aligned with the radio synchrotron jet-emission have been
reported by Kadler et al. (2004); they derive for the
fitting of the core spectrum $\Gamma\approx$0.25 and kT$\approx$0.5~keV.
The value they estimate for the luminosity is within the factor of 3 of
our estimation (see Table \ref{tab:lumflux}).
\smallskip

{\bf NGC\,2681 (UGC\,4645).}  An unresolved nuclear source is clearly
detected at hard X-ray energies (Fig. \ref{fig:clasif}). Satyapal et al. (2005), who made
use of archival \emph{Chandra} ACIS observations of this galaxy, classed by
them as an AGN-LINER, and derived kT=0.73~keV and $\Gamma$=1.57 for a
apec plus power-law fit to the nuclear spectrum. These values are in
perfect agreement, within the errors, with the parameters we derive
for our best model (ME+PL) ($\Gamma$=1.74 and kT=0.66~keV, see Table 
\ref{tab:fittings}), that we recall gives unacceptable values of $\chi^{2}$.
\smallskip

{\bf UGC\,05101.} In addition to the hard band point nuclear source,
extended emission is seen in both (4.5-8.0$^*$~keV) and (6-7~keV) bands (Fig. \ref{fig:clasif}).
The evidence of a buried active galactic nucleus in this ultra-luminous
infrared galaxy has been provided by the analysis by Imanishi et
al. (2003) of its \emph{XMM-Newton} EPIC spectrum. They fit the spectrum with
an absorbed power-law ($\Gamma$=1.8 fixed), a narrow gaussian for
the 6.4~keV Fe K$\alpha$ line, which is clearly seen in their
spectrum, and a 0.7~keV thermal component, deriving (N$\rm{_H}$=14
$\times$ 10$^{23}$cm$^{-2}$) and EW(Fe K$\alpha$)=0.41~keV. The
resulting (2-10~keV) luminosity of $\sim$5$\times$10$^{42}$erg~s$^{-1}$ 
is about 30 times higher than
the value we
estimate. Fe-K emission is marginally detected in the analysis of
\emph{Chandra} data by Ptak et al. (2003). The luminosity we estimate is a
factor of two higher than that given by Dudik et al. (2005) derived
from \emph{Chandra} ACIS data. 
\smallskip

{\bf NGC\,3226 (UGC\,5617, Arp\,94A).} Several point sources are detected
at the (4-8)~keV band image of this galaxy, with Fe emission
unambiguously present in the nucleus (Fig. \ref{fig:clasif}).  The analysis of HETGS
\emph{Chandra} data by George et al. (2001), whose properties strongly
suggested that this galaxy hosted a central AGN, resulted in an
adequate fit with a photon index $\Gamma$=1.94 and N$\rm{_H}$=4.8 $\times$
10$^{21}$cm$^{-2}$, with the implied luminosity L(2-10~keV)$\approx$3.2$\times$10$^{40}$ 
erg~s$^{-1}$.
\emph{XMM-Newton} observations of this dwarf elliptical most probably
indicate the presence of a sub-Eddington, super-massive black hole in a
radiative inefficient stage (Gondoin et al. 2004). They conclude
that, since the best fit is provided by a bremsstrahlung model
absorbed by neutral material, the X-ray emission may therefore be
reminiscent of advection-dominated accretion flows. Nevertheless, an
acceptable fit is also obtained by including a power-law model
($\Gamma$=1.96) absorbed by neutral (N$\rm{_H}$=4.1
$\times$
10$^{21}$cm$^{-2}$) and ionized material. The resulting (2-10~keV)
luminosity, calculated for the distance we use, is 1.8 $\times$
10$^{40}$ erg~s$^{-1}$, a factor of 4 higher than the one we estimate.
Terashima \& Wilson (2003) fit the \emph{Chandra} ACIS nuclear spectrum with
a power-law with $\Gamma$=2.21 (from 1.62 to 2.76) and N$\rm{_H}$=0.93
$\times$ 10$^{22}$cm$^{-2}$. Notice that substantial absorption is
also derived from the position of this galaxy in the color-color
diagrams, whereas the power-law index we estimate is somewhat steeper
 (see Table 6).
\smallskip

{\bf NGC\,3245 (UGC\,5663).}  No unresolved nuclear source is detected
in the (4.5-8.0$^*$~keV) band image. This contrasts with the analysis
by Filho et al. (2004) who, making use of the same \emph{Chandra}
data, conclude that there is a hard nuclear X-ray source coincident
with the optical nucleus. The luminosity they calculate with a fixed
$\Gamma$=1.7 is less than a factor of 2 fainter than ours.
\smallskip

{\bf NGC\,3379 (UGC\,5902, M\,105).}  
Very recently, David et al. (2005) have published their study
of the X-ray emission as traced by ACIS-S \emph{Chandra} observations, which
is mainly devoted to the analysis of extra-nuclear X-ray sources and
diffuse emission, and they derive a power-law index for the diffuse
emission of 1.6-1.7, in agreement with the value reported by
Georgantopoulos et al. (2002). David et al. (2005) do not fit the
spectrum of the nuclear source (their source 1) due to the too few net
counts in the S3 chip data for this object. This is also the reason for 
not having neither a fit nor an estimation of the spectral parameters
(see Table 6).
\smallskip

{\bf NGC\,3507 (UGC\,6123).}  No hard nuclear point source is detected
for this galaxy (Fig. \ref{fig:clasif}). The only previous published X-ray study is
based on observations obtained with \emph{ASCA}: Terashima et al. (2002) get
$\Gamma$=2.3 and N$\rm{_H}$=1.0 $\times$ 10$^{22}$cm$^{-2}$, but conclude
that a power-law model ($\Gamma$=1.7) describes the spectrum well.
\smallskip

{\bf NGC\,3607 (UGC\,6297).} No hard nuclear point source is detected
for this galaxy (Fig. \ref{fig:clasif}).  Based on observations obtained with \emph{ASCA},
Terashima et al. (2002) found no clear evidence for the presence of an
AGN in this LINER, in agreement with our classification.
\smallskip

{\bf NGC\,3608 (UGC\,6299).}  No hard nuclear point source is detected
for this galaxy (Fig. \ref{fig:clasif}). The only previous X-ray study of this galaxy
is that by O'Sullivan et al.  (2001) who present a catalogue of X-ray
bolometric luminosities for 401 early-type galaxies obtained with
\emph{ROSAT} PSPC pointed observations. Corrected to our adopted distance,
this luminosity is of 1.37 $\times$ 10$^{40}$erg~s$^{-1}$, about 30
times brighter than our estimation.
\smallskip

{\bf NGC\,3628 (UGC\,6350, Arp\,317C).} The hard X-ray morphology shows
an unresolved nuclear component that also appears in the Fe image
(Fig. \ref{fig:clasif}).  \emph{Chandra} X-ray and ground-based optical H$\alpha$, arc-second
resolution imaging is studied by Strickland et al. (2004), with the
main aim of determining both spectral and spatial properties of the
diffuse X-ray emission. They also show the total counts for the
nuclear region (an extraction of 1 kpc radius around the dynamical
center that, for this galaxy, corresponds to the central 20''), 
but no spectral fitting is attempted.  Our morphological
classification does not agree with that of Dudik et al. (2005) who have
classified this galaxy as an object displaying no nuclear source
according to its morphology in \emph{Chandra} ACIS data; this galaxy is taken
as a LINER/transition object and an upper limit of 2.7 $\times$
10$^{37}$erg~s$^{-1}$ (corrected to our adopted distance) is given for
its (2-10~keV) nuclear luminosity, about 6 times fainter than our
estimated luminosity. Note that a high absorption is derived from the
position of this galaxy in the color-color diagrams.
\smallskip

{\bf NGC\,3690B (Arp\,299, Mrk\,171).}  X-ray emission is plenty of
features, with a hard unresolved source clearly detected in the
nuclear position, which is also seen in the 6-7~keV band (Fig. \ref{fig:clasif}). EPIC-pn
\emph{XMM-Newton} spatially resolved data have clearly demonstrated the
existence of an AGN in NGC\,3690, for which a strong 6.4~keV line is
detected, and suggested that the nucleus of its companion IC\,694 might also host an
AGN\footnote{Note however that both galaxies are found in the
comparative sample of starburst galaxies in Satyapal et al. (2004).},
since a strong 6.7~keV Fe-K$\alpha$ line is present (Ballo et
al. 2004). Considering both that \emph{XMM-Newton} integrates a larger area
than the nuclear 3'' we extract and that a gaussian line
together with a single power-law fit ($\Gamma$=1.8 and N$\rm{_H}$=5.6
$\times$ 10$^{21}$cm$^{-2}$) is used for the fitting of EPIC-pn data,
the differing results can be explained.
\smallskip

{\bf NGC\,4111 (UGC\,7103).} A hard nuclear point source is detected for
this galaxy (Fig. \ref{fig:clasif}). Previous X-ray spectral analysis comes from \emph{ASCA}
data by Terashima et al. (2000,  see also Terashima et
al. 2002) who could not fit the spectrum with a single component
model, but a combination of a power-law together with a Raymond-Smith
plasma, with $\Gamma$=0.9, kT=0.65~keV (in reasonable agreement with
the parameters we estimate from its position in the color-color
diagrams, see Table 4) and N$\rm{_H}$=1.4 $\times$ 10$^{20}$cm$^{-2}$
reproduced well the observed spectrum and gives an intrinsic L(2-10~keV)=
6.8$\times$ 10$^{39}$ erg~s$^{-1}$, a factor of 3 higher than the
one we estimate (see Table \ref{tab:lumflux}).
\smallskip

{\bf NGC\,4125 (UGC\,7118).}  Fig. \ref{fig:clasif} shows the presence
of a nuclear hard point source. The best fit Georgantopoulos et
al. (2002) obtain for the central 2\arcmin~ \emph{BeppoSAX} spectrum
is that provided by an absorbed power-law with $\Gamma$=2.52 and
N$\rm{_H}$=3$\times$ 10$^{22}$cm$^{-2}$, providing
L(2-10~keV)=0.68$\times$ 10$^{40}$ erg~s$^{-1}$. Satyapal et
al. (2004), based on \emph{Chandra} ACIS imaging, class this galaxy
among those revealing a hard nuclear source embedded in soft diffuse
emission; they estimate the luminosity by assuming an intrinsic
power-law slope of 1.8 which results to be (once corrected to our
adopted distance) L(2-10~keV)=7.3$\times$ 10$^{38}$ erg~s$^{-1}$, in
very good agreement with the value we estimate.
\smallskip

{\bf NGC\,4261 (UGC\,7360, 3C\,270).} The nuclear hard band emission of
this galaxy is clearly unresolved both in the (4.5-8.0$^*$~keV) and 6-7~keV bands
(Fig. \ref{fig:clasif}). Sambruna (2003) published its nuclear EPIC-pn
\emph{XMM-Newton} spectrum (the central 10'') which is best fitted
with a two component model with a power-law ($\Gamma$=1.4) absorbed by
a column density of N$\rm{_H}$$\approx$4$\times$ 10$^{22}$cm$^{-2}$ plus a
thermal component with kT$\approx$0.7~keV (in agreement with \emph{Chandra}
spectral results by Gliozzi et al. 2003 and Chiaberge et al. 2003); an
unresolved FeK emission line with EW$\approx$0.28~keV is detected at
$\sim$7~keV. They also report short-term flux variability from the
nucleus (timescale of 3-5 ks), that they argue as being originated in
the inner jet. The various features seen at soft energies (Fig. \ref{fig:clasif})
were already shown by Donato et al. (2004), who also analyzed its
\emph{Chandra} ACIS data for the core component (core radius of 0.98''), 
that they fit with a PL+apec model with $\Gamma$=1.09,
kT=0.60~keV and a high column density N$\rm{_H}$=7.0$\times$
10$^{22}$cm$^{-2}$, reported to be the largest intrinsic column
density of the 25 radio galaxies in their study. These parameters
agree with those obtained by Rinn et al. (2005) and Satyapal et
al. (2005) for the same data.

Very recently Zezas et al. (2005) have published the analysis of 35ks
\emph{Chandra} ACIS-S observations. They report an almost point-like emission
above 4.0~keV, and evidence for an X-ray jet component down to
arc-second scales from the nucleus (barely visible in our Fig. \ref{fig:clasif}). A
three component model is given as the best fit for the X-ray spectrum
of the nuclear 2'': a heavily obscured, flat power-law
($\Gamma$=1.54 and N$\rm{_H}$=8.4$\times$ 10$^{22}$cm$^{-2}$), a less
absorbed, steeper power-law ($\Gamma$=2.25 and N$\rm{_H}$$<$ 3.7$\times$
10$^{20}$cm$^{-2}$) and a thermal component (kT=0.50~keV), which
results in L(2-10~keV)=10.8$\times$ 10$^{40}$ erg~s$^{-1}$, a factor of
2 higher that what we estimate. They report an equally good fit with a
single power-law ($\Gamma$=1.37) seen through a partially covering
absorber (N$\rm{_H}$=7.7$\times$ 10$^{22}$cm$^{-2}$, f$_{cov}$=0.92) plus a
thermal component. We have not included this object in the SF subsample 
due to its complexity that gives as but fitted and unexpected parameters with 
our five models. 
\smallskip

{\bf NGC\,4314 (UGC\,7443).} No nuclear source is detected in the hard
X-ray band (Fig. \ref{fig:clasif}).  Satyapal et al. (2004) use \emph{Chandra} ACIS imaging
to classify this galaxy among those exhibit multiple, hard off-nuclear
point sources of comparable brightness to the nuclear source; with an
assumed power-law index of 1.8, the corresponding luminosity,
corrected to our adopted distance, results to be L(2-10~keV)=8$\times$ 
10$^{37}$ erg~s$^{-1}$, in excellent agreement with the one we estimate 
(see Table \ref{tab:lumflux}).
\smallskip

{\bf NGC\,4374 (M\,84, UGC\,7494, 3C\,272.1).} Both in (4.5-8.0$^*$~keV) and
6-7~keV band images an unresolved nuclear source is detected (Fig. \ref{fig:clasif}).
Satyapal et al. (2004) already described the X-ray morphology traced
by \emph{Chandra} ACIS imaging of this galaxy as revealing a hard nuclear
source embedded in soft diffuse emission. The \emph{Chandra} ACIS-S data are
analyzed by Finoguenov \& Jones (2001)\footnote{See also Kataoka \&
Stawarz (2005) for the analysis of the two extra-nuclear
knots.}, who report a remarkable interaction of the radio lobes and
the diffuse X-ray emission, and provide the parameters for a fit with
an absorbed (N$\rm{_H}$=2.7$\times$ 10$^{21}$cm$^{-2}$) power-law
($\Gamma$=2.3) and the corresponding L(0.5-10~keV)=4.7$\times$
10$^{39}$ erg~s$^{-1}$, all in very good agreement with the ones we
give in this paper (see Tables \ref{tab:fittings_anex} and \ref{tab:fittings});
 but they somewhat differ from
the ones obtained from the \emph{ASCA} spectrum (Terashima et al. 2002), most
probably due to the different spatial resolutions.
\smallskip

{\bf NGC\,4395 (UGC\,7524).}  The unresolved nuclear source is seen in
both (4.5-8$^*$~keV) and 6-7~keV band images (Fig. \ref{fig:clasif}). Moran et
al. (2005) have recently published\footnote{See also O'Neill et al. 2006, ApJ accepted, astro-ph/0603312.} the first high-quality, broadband
X-ray detection of the AGN of this object, confirming the rapid, large
amplitude variability reported in previous studies (Iwasawa et
al. 2000, Shih et al. 2003) and confirmed with \emph{XMM-Newton} EPIC-pn data
(Vaughan et al. 2005). They fit a single power-law model with
absorption by neutral material to the spectrum of the nuclear 5'' 
($\Gamma$=0.61 and N$\rm{_H}$=1.2$\times$ 10$^{22}$cm$^{-2}$),
that provides a poor fit over the entire \emph{Chandra} 0.5 to 9~keV range,
but they claim to be excellent for energies above $\sim$1.2~keV. 
Our best model results to be a single absorbed power-law as well,
but with somewhat stronger absorption (N$\rm{_H}$=2.87$\times$
10$^{22}$cm$^{-2}$) and a higher spectral index ($\Gamma$=1.44, see
Table 6). The iron line clearly resolved in the time-averaged \emph{ASCA}
spectrum shown by Shih et al. (2003) is also visible in our spectrum
(see Fig. \ref{fig:espectroNGC6240}).
\smallskip

{\bf NGC\,4410A (UGC\,7535, Mrk\,1325).} Both (4.5-8$^*$~keV) and 6-7~keV band
images show the unresolved nature of the nuclear source at these
energies (Fig. \ref{fig:clasif}).  ACIS-S \emph{Chandra} observations of the NGC\,4410 group
are presented in Smith et al. (2003), who obtained an adequate fit for
the spectrum of the inner 1'' to a power-law with
$\Gamma$$\approx$2 and a fixed N$\rm{_H}$=5$\times$ 10$^{20}$cm$^{-2}$,
in agreement with previous analysis of \emph{ROSAT} X-ray observations
(Tsch\"oke et al. 1999). Our best model only needs the inclusion of a power
law with $\Gamma$=1.75 (consistent with theirs within the errors) 
\smallskip

{\bf NGC\,4438 (UGC\,7574, Arp\,120B).} No nuclear point source is
detected in our hard X-ray band images (Fig. \ref{fig:clasif}). The results from
25~ks \emph{Chandra} ACIS-S observations of this galaxy are presented in
Machacek et al. (2004) who, in contrast with our morphological
classification, suggest the presence of an AGN based on the steep
spectral index and the location of the hard emission at the center of
the galaxy. The spectrum of the central 5'' is claimed to be
best fitted by a combination of an absorbed power-law (with N$\rm{_H}$=2.9$\times$ 
10$^{22}$cm$^{-2}$ and a fixed $\Gamma$=2.0) and a MEKAL
0.58~keV thermal component, providing L(2-10~keV)=2.5$\times$
10$^{39}$ erg~s$^{-1}$. Nevertheless,
Satyapal et al. (2005) class this galaxy as an non AGN-LINER based on
its ACIS \emph{Chandra} image, in agreement with our classification, with
kT=0.77~keV and N$\rm{_H}$=1.2$\times$ 10$^{21}$cm$^{-2}$, consistent 
with ours within the errors, but $\Gamma$=1.19.
\smallskip

{\bf NGC\,4457 (UGC\,7609).}  Hard emission is unresolved in the nucleus
of this galaxy (Fig. \ref{fig:clasif}). The spectral analysis of ACIS \emph{Chandra} data by
Satyapal et al. (2005) gives $\Gamma$=1.57, kT=0.69~keV and no additional
absorption, in very good agreement with our results (see Table 6).
\smallskip

{\bf NGC\,4459 (UGC\,7614).} 
Our morphological classification (SB candidate, see Fig. \ref{fig:clasif}) agrees
with that by Satyapal et al. (2005), also based on ACIS \emph{Chandra} data,
who gives no additional X-rays information on this object.
\smallskip

{\bf NGC\,4486 (M\,87, UGC\,7654, Virgo\,A, Arp\,152, 3C\,274).}  Both the
unresolved nuclear emission and the jet-like feature extending
$\sim$15'' to the W-NW, in the direction of the optical
jet, are seen in Fig. \ref{fig:clasif}. Combined deep \emph{Chandra}, \emph{ROSAT} HRI and
\emph{XMM-Newton} observations of this galaxy are shown in Forman et
al. (2005), where the same salient features present in our Fig. \ref{fig:clasif} can
be seen, with X-ray jets clearly detected, but no spectral analysis is
made. Donato et al. (2005) analyze both \emph{Chandra} and \emph{XMM-Newton} data
providing a radius for the core of 0.22''. Dudik et
al. (2005) classed it among objects exhibiting a dominant hard
nuclear point source and estimate its luminosity as
L(2-10~keV)=3.3$\times$ 10$^{40}$ erg~s$^{-1}$ with a fixed
$\Gamma$=1.8 power-law, in good agreement with the one we estimate
(see Table \ref{tab:lumflux}).
\smallskip

{\bf NGC\,4494 (UGC\,7662).}  Hard nuclear emission is point-like
(Fig. \ref{fig:clasif}). The \emph{XMM-Newton} EPIC spectrum extracted from a 45''
region has been published by O'Sullivan \& Ponman (2004). In agreement
with our findings, a ME+PL combination results to be the best model
for the spectral fitting, for which they get $\Gamma$=1.5
(consistent with our value) but for
hydrogen column density fixed at the Galactic value (N$\rm{_H}$=1.56$\times$ 
10$^{20}$cm$^{-2}$  and
kT=0.25~keV). In agreement with our morphological classification, Dudik et
al. (2005) class it as a hard nuclear point dominated source and
estimate L(2-10~keV)=7.2$\times$ 10$^{38}$ erg~s$^{-1}$ with a fixed
$\Gamma$=1.8 power-law, about a factor of 6 fainter 
than the one we calculate with the
spectral fitting.
\smallskip

{\bf NGC\,4552 (M\,89, UGC\,7760).}
This galaxy shows in the hard band an unresolved source over an extended 
nebulosity  morphology with the peak of 
emission coincident with the galaxy center determined from 2MASS data (Fig. \ref{fig:clasif}).
Xu et al. (2005) found from \emph{Chandra} ACIS-S data that the central source is the 
brightest in the field and it coincides with the optical/IR/radio center of
the galaxy within 0.5''. The X-ray identified source is compact and 
variable in short time scales of 1 h. Their best fitted model to the source
is consistent with an absorbed power-law with spectral index $\Gamma$=2.24
in rather good agreement with the \emph{ASCA} data reported 
by Colbert and Mushotzky (1999). The inferred luminosity in the 2-10~keV 
was of 4 $\times$ 10$^{39}$ erg~s$^{-1}$, consistent with our result 
(2.6$\times$ 10$^{39}$erg s$^{-1}$). Their main conclusion based 
on the found variability, the spectral analysis and multi-wavelengh data is 
that the central source is most likely a low luminosity AGN than contribution 
from LMXBs.  
Our best fit parameters are consistent with a model of a power 
law with an spectral index $\Gamma$=1.81
plus a thermal RS 
of kT= 0.83,
in much better agreement with the results by Filho et al. (2004) on the 
analysis of \emph{Chandra} archival data, with $\Gamma$=1.51 and kT=0.95.
\smallskip

{\bf NGC\,4579 (M\,58, UGC\,7796).}  This galaxy shows a compact nuclear
source sitting in a diffuse halo (Fig. \ref{fig:clasif}). Eracleous et al. (2002)
fitted the compact unresolved central source detected in \emph{Chandra} X-ray
data, coincident with the broad line region detected in UV by Barth et
al. (2002) with a simple power-law spectra with $\Gamma$=1.88
which gives an estimated luminosity of 1.7$\times$ 10$^{41}$erg
s$^{-1}$. More recently Dewangan et al. (2004) present \emph{XMM-Newton} data
to search for the presence of FeK$\alpha$ line. The best fit spectrum
is rather complex: an absorbed power-law with $\Gamma$=1.77
plus a narrow gaussian at 6.4~keV and a broad gaussian at 6.79~keV
with FWHM $\sim$ 20.000 km s$^{-1}$. This broad component is
interpreted as arising from the inner accretion disk. The estimated
luminosity amounts to 1.2$\times$ 10$^{40}$erg s$^{-1}$, being lower
than both Eracleous's estimation and ours (1.4$\times$ 10$^{41}$erg s$^{-1}$).
\smallskip

{\bf NGC\,4594 (M\,104, Sombrero Galaxy).}  The Sombrero galaxy shows
the typical X-ray morphology of a compact unresolved nuclear source on
top of a diffuse halo (Fig. \ref{fig:clasif}). Dudik et
al. (2005) class it within the objects that exhibit a dominant hard
nuclear point source. Pellegrini et al. (2003) present an
investigation with \emph{XMM-Newton} and \emph{Chandra} of the 7''
central nuclear source being consistent with an absorbed power-law of
$\Gamma$=1.89
with a column density of N$\rm{_H}$=1.8$\times$ 10$^{21}$cm$^{-2}$, in
close agreement, within the errors, with our fitted values. Our value
of the estimated 2-10~keV luminosity, 1.2$\times$ 10$^{40}$
erg~s$^{-1}$, agrees pretty well with the data reported by Pellegrini
from \emph{XMM-Newton}. 
\smallskip

{\bf NGC\,4596 (UGC\,7828).}  This galaxy is very faint at X-ray
frequencies, 
showing diffuse X-ray morphology in all the spectral bands
(Fig. \ref{fig:clasif}). Information on its spectral properties cannot be obtained
based on the present data due to the lack of sufficient counts in the
hard band (4.5-8.0$^{*}$~keV). No previous X-ray data have been reported for this
galaxy.
\smallskip

{\bf NGC\,4636 (UGC\,7878).}  This galaxy does not show emission at high
energies (Fig. \ref{fig:clasif}). \emph{Chandra} data do not have enough quality to allow a
proper fitting to the spectrum. Xu et al. (2002) and O'Sullivan et al
(2005) present \emph{XMM-Newton} data for this source and obtain that it can
be consistent with thermal plasma with a temperature kT between 0.53
and 0.71~keV. The arm-like structure reported by Jones et al. (2002) at
soft energies can be produced by shocks driven by symmetric off-center
outbursts preventing the deposition of gas in the center. O'Sullivan
et al. (2005) suggest that the X-ray morphology can be the result of a
past AGN being quiescent at the present. The difference in our
luminosity estimation (1.77 $\times$ 10$^{39}$erg s$^{-1}$) and the
value reported by Loewenstein et al. (2001) for the nucleus (2
$\times$ 10$^{38}$erg s$^{-1}$) is due to the different apertures used,
13'' and  3'', respectively.
\smallskip

{\bf NGC\,4676A and B (Arp\,242, The Mice Galaxy).}  
No detection of X-rays can be seen
at high energies (Fig. \ref{fig:clasif}). Read (2003) present the first \emph{Chandra}
analysis of The Mice Galaxy and found a compact source in component B being
rather diffuse in A. Their spectral fitting in B is both consistent with
MEKAL thermal and power-law modeling. We did not perform any fitting
due to poor counting statistics. From the color-color diagrams the data
for component A are consistent with a power-law with an spectral
index in the range 0.8-1.2. For component B we did not make any
estimation since the error in the count rate for the hardest band is
greatest than 80\%.  Our estimated luminosities for both components
are in remarkably good agreement with the results by Read and can be
attributed to the starbursts in both components.
\smallskip

{\bf NGC\,4696 (Abell\,3526).}  This galaxy is rather diffuse at high
energies, having a clear nuclear halo morphology at soft energies
(Fig. \ref{fig:clasif}). In fact, Dudik et al. (2005) class it as an object that
reveals a hard nuclear point source embedded in soft diffuse
emission. Di Matteo et al. (2000) and Allen et al. (2000) in their
reported \emph{ASCA} data for a few giant ellipticals in clusters, 
include this galaxy which is the center of the Centaurus cluster, the
analysis of its spectrum showing a high luminosity of 2.14$\times$
10$^{42}$erg s$^{-1}$.  Taylor et al. (2006) obtain the best fit model
by a MEKAL thermal plasma with kT=0.75~keV 
and abundance of 0.22 times the solar abundance; in the same
sense, Rinn et al. (2005) fit its \emph{XMM-Newton} spectrum with a thermal
model with kT=0.7~keV but for a 1.2 solar metallicity. At variance with
them, our best fit model seems to be a power-law but with rather high
and unrealistic spectral index of 4.26.
This difference can be attributed to the different aperture used, 3.9''
in our case and 0.9'' in the data by Taylor et al. Even so, the
estimated luminosities are not very far within a factor of 2
(6$\times$ 10$^{39}$erg s$^{-1}$ and 1.2$\times$ 10$^{40}$erg s$^{-1}$
for our analysis and Taylor's, respectively). We have classified this
source as a good candidate to be an Starburst due to the absence of a
nuclear unresolved source at hard energies (Fig. \ref{fig:clasif}). Nevertheless, The
VLBA data reported by Taylor and collaborators reveal a weak nucleus and a broad,
one-sided jet extending over 25 pc suggesting the AGN nature of this
peculiar source.
\smallskip
 
{\bf NGC\,4698 (UGC\,7970).}  This galaxy shows very faint high energy
X-ray emission on its central region.  The largest extension is found
at intermediate energies, between 1 and 4~keV (Fig. \ref{fig:clasif}).
Georgantopoulos \& Zezas (2003) make a careful analysis of the
\emph{Chandra} data on this source and found that the X-ray nuclear position
coincide with the faint radio source reported
by Ho and Ulvestad (2001). They found that the best fit model consist
of an absorbed power-law with $\Gamma$=2.18
and column density of N$\rm{_H}$=5 $\times$ 10$^{20}$cm$^{-2}$ which
gives a nuclear luminosity of 10$^{39}$erg s$^{-1}$.  We found from
the color-color diagrams that the data are consistent with a combined
model of a power-law with $\Gamma$=[1.2-1.6] and a thermal component
with kT=[0.7-0.8]~keV and a luminosity lower by a factor of two than
the one estimated by Georgantatopoulos \& Zezas (2003). Cappi et
al. (2006) fit its \emph{XMM-Newton} spectrum with a single power-law model
with $\Gamma$=2.0 and get L(2-10 keV)=1.6$\times$ 10$^{39}$erg
s$^{-1}$, a factor of 3 brighter than our determination.
\smallskip

{\bf NGC\,4736 (M\,94, UGC\,7996).} This galaxy shows a large amount of
unresolved compact sources in the central few arc-sec, which make the
extraction of the true nuclear source rather difficult
(Fig. \ref{fig:clasif}). Eracleous et al. (2002) identified 3 sources in the nuclear
region, all of them showing hard spectra with power-law indices
ranging from 1.13 for the brightest one till 1.8 for X-3, and
luminosities in the 2-10~keV band between 4$\times$ 10$^{38}$ erg~s$^{-1}$
and 9.1 $\times$ 10$^{39}$erg s$^{-1}$. We have identified the source X-2 by Eracleous 
as the nucleus of the galaxy since it coincides with the 2MASS
near-IR nucleus within 0.82''. 
Eracleous (2002) pointed out on the complications
to define an AGN or SB character to this source suggesting that even
if the brightest source is associated with an AGN it only will
contribute in a 20\% to the energy balance in the X-rays.  The radio
monitoring observations made by K\"ording et al. (2005) with the VLBI
found a double structure, the radio position N4736-b coinciding with
our identified X-ray nucleus. From this double structure the brightest
knot N4736-b appears to be also variable, pointing to an AGN nature of
this low luminosity AGN.
\smallskip

{\bf NGC\,5055 (M\,63, UGC\,8334).}  This galaxy shows a clear unresolved
nuclear source coincident with the 2MASS position for the nucleus
(Fig. \ref{fig:clasif}). No previous \emph{Chandra} data have been reported. The only data
available were \emph{ROSAT} PSPC and HRI observations (Read et al. 1997,
Roberts and Warwick 2000) pointing to the nucleated nature of this
source within the low spatial resolution, 10'' at best. Recently, 
Liu and Bregman (2005) in the course of an investigation of ULX over a
sample of 313 nearby galaxies found 10~ULX in this galaxy, one of these
being close to the nucleus with a variable luminosity between 0.96
and 1.59 $\times$ 10$^{39}$erg s$^{-1}$ in 1.6 days. 
\smallskip

{\bf NGC\,5194 (M\,51a, UGC\,8493, Arp\,85a).}  A clear unresolved nuclear
source is identified in the hard band of M51 coincident within 2.87"
with the near IR nucleus (Fig. \ref{fig:clasif}). Dudik et al. (2005) class it
as an object that reveal a hard nuclear point source embedded in soft
diffuse emission. Its spectral properties suggest
that the source can be modeled by a combination of MEKAL at
kT=0.61~keV
plus a power-law with
$\Gamma$=2.67
and column density consistent with the
galactic value, this fitting providing 
a fairly low luminosity in the hard band (1.38 $\times$ 10$^{38}$erg
s$^{-1}$). Dewangan et al. (2005) obtain \emph{XMM-Newton}
observations for the galaxy that show an extremely flat continuum and
a narrow iron K$\alpha$ line. They investigate different models for
the galaxy, the best one being more consistent with a reflection of
the primary power-law ( $\Gamma$=1.9) by cold and dense material. By
using this model they estimate a luminosity of 1.8$\sim$ 10$^{39}$erg
s$^{-1}$, which is a factor of 10 larger than our estimation. 
Cappi et
al. (2006) fit its \emph{XMM-Newton} spectrum with a combined power-law 
with $\Gamma$=0.6 and thermal with kT=0.5~keV, together with a Fe K line 
with EW(Fe K)=0.986~keV, and get 
L(2-10 keV)=3.3$\times$ 10$^{39}$erg
s$^{-1}$, a factor of 20 brighter than our determination. These
differences can be attributed either to the different model used or
maybe to the different spatial resolution of \emph{XMM-Newton} and
\emph{Chandra} data. It has to be noticed that the iron line FeK has not been
included in our fitting but it is clearly detected. We note that whereas 
UGC\,08696 shows a compact nuclear
source in this energy band it can not be directly associated with a
FeK line because it has a broad high energy component at these energies.
\smallskip

{\bf Mrk\,266 (NGC\,5256, UGC\,8632, IZw\,67).}  Its X-ray morphology
shows the double structure of these merging system with the North-West
nucleus being brighter than the southern one. Also the southern
nucleus shows hard emission being more diffuse (Fig. \ref{fig:clasif}).  
New \emph{XMM-Newton} observations have been reported by Guainazzi et
al. (2005), with data consistent with a thermal plasma of luminosity
3.6$\times$ 10$^{40}$erg s$^{-1}$, which seems to be in good agreement
with our results (see Table \ref{tab:lumflux}).
\smallskip

{\bf UGC\,08696 (Mrk\,273).}  Mrk\,273 is one of the prototypical
Ultra-luminous Galaxies showing a very complex structure at optical
frequencies with a double nuclei and a long tidal tail. At high X-ray
energies only the northern nucleus is detected (Fig. \ref{fig:clasif})
which it is coincident with the compact radio source shown by VLBI
observations (Cole et al. 1999, Carilli and Taylor 2000). Satyapal et
al. (2004), based on \emph{Chandra} ACIS imaging, class this galaxy
among those revealing a hard nuclear source embedded in soft diffuse
emission. Xia et
al. (2002) report previous analysis of the X-ray \emph{Chandra} data;
from their careful analysis of both the nucleus and the extended
emission, they show that the compact nucleus is well described by an
absorbed power-law (N$\rm{_H}$= 4.1$\times$10$^{20}$cm$^{-2}$,
$\Gamma$=2.1, L(2-10~keV)=2.9$\times$ 10$^{42}$ erg~s$^{-1}$) plus a
narrow FeK$\alpha$ line. The most remarkably result from this analysis
is that the spectrum of the central 10" is consistent with a
metallicity of 1.5Z$_\odot$ whereas the extended halo seems to be
consistent with a thermal plasma with metallicity of 0.1 Z$_\odot$.
The results reported by Ptak et al. (2003) pointed out that most of
the observed X-ray emission (95\%) comes from the nucleus.  Balestra
et al. (2005) using \emph{XMM-Newton} data analyze the FeK$\alpha$
line and conclude that, as in the case of NGC\,6240, the line is the
result of the superposition of neutral FeK$\alpha$ and a blend of
highly ionized lines of FeXXV and FeXXVI.  Our best fit model is in
good agreement with these data within the errors ($\Gamma$=1.74
, kT=0.75~keV
, N$\rm{_H}$=3.9 $\times$10$^{20}$cm$^{-2}$ and
L(2-10~keV)=1.5 $\times$ 10$^{42}$ erg~s$^{-1}$, see Tables
\ref{tab:lumflux} and \ref{tab:fittings}).
\smallskip

{\bf CGCG\,162-010 (Abell\,1795, 4C\,26.42).}  This galaxy is the
central galaxy of the cluster A1795, which hosts the powerful type I
radio source 4C26.42. The X-ray morphology shows a rather diffuse
emission at high energies and very clear long filament at soft
energies (Fig. \ref{fig:clasif}). A full description of the nature of
this filament is made in Crawford et al. (2005), who attribute the
observed structure as due to a large event of star formation induced
by the interaction of the radio jet with the intra-cluster
medium. Satyapal et al. (2004), based on \emph{Chandra} ACIS imaging,
class this galaxy among those revealing a hard nuclear source embedded
in soft diffuse emission. Nevertheless, Donato et al. (2004)
investigate the nature of the X-ray central compact core in a sample
of type I radio galaxies and classified this galaxy among sources
without a detected compact core, in agreement with our classification.
Our X-ray spectroscopic analysis
results in this object being one of the five most luminous in our
sample, with a value in very good agreement with the one
estimated by Satyapal et al. (2004) for an intrinsic power-law slope
of 1.8 (see Table \ref{tab:lumflux}).
\smallskip

{\bf NGC\,5746 (UGC\,9499).}
No previous X-ray data have been previously reported. Its morphology shows
a clear compact unresolved nuclear source (Fig. \ref{fig:clasif}).
Nagar et al. (2002) detected a compact radio source suggesting the AGN
nature of this galaxy. Both the fitting and the position in the
color-color diagrams indicate considerable obscuration (see Table 6).
\smallskip

{\bf NGC\,5846 (UGC\,9706).}  Trinchieri and Goudfroij (2002), based on
\emph{Chandra} data, reveal a complex X-ray morphology with no clear nuclear
identification (see also Fig. \ref{fig:clasif}). They detect, however,
a large amount of individual compact sources in the range of
luminosities of 3-20$\times$ 10$^{38}$ erg s$^{-1}$.
Filho et al. (2004) report a weak hard (2-10 keV) nuclear source
with $\Gamma$=2.29, compatible within the errors with the value we
obtain from the spectral fitting. Satyapal et al. (2005) analyze the
Chandra data of this galaxy that they class within non-AGN LINERs,
fitting its spectrum with a single thermal model with kT=0.65 keV,
exactly the same we get for our single RS model (see Table 3).
\smallskip
 
{\bf NGC\,5866 (UGC\,9723).}  The data for this galaxy reveal a rather
complex morphology at hard frequencies with an identifiable nuclear
region and extended emission in the North-West direction
(Fig. \ref{fig:clasif}). Previous X-ray data by Pellegrini (1994) are
based on \emph{ROSAT} PSPC observations, where they pointed out the
high excess of soft X-ray emission in S0 galaxies. Filho et
al. (2004) and Terashima \& Wilson (2003) failed to detect any hard
nuclear X-ray emission in the \emph{Chandra} image of this galaxy, and
Satyapal et al. (2005) class it as a non-AGN-LINER, what agrees with
our morphological classification.
\smallskip

{\bf NGC\,6251 (UGC\,10501).}  This is a well known radio galaxy hosting
a giant radio jet (Birkinshaw and Worrall 1993, Sudou et
al. 2001). The high energy X-ray morphology shows a well defined
unresolved nuclear source without any extended halo (Fig. \ref{fig:clasif}).
Guainazzi et al. (2003) reported a full analysis of the nuclear energy
source comparing \emph{Chandra}, \emph{BeppoSAX} and \emph{ASCA} data. They found that the
spectra can be modeled by a combination of thermal plasma at
kT=1.4~keV
plus a power-law with $\Gamma$=1.76
and N$\rm{_H}$=1.6$\times$ 10$^{21}$cm$^{-2}$, but they do not find
evidences of the broad FeK$\alpha$ claimed by previous \emph{ASCA}
observations. However, the high sensitivity of \emph{XMM-Newton} leads
Gliozzi et al. (2004) to suggest again that such a broad ($\sigma$=
0.6~keV) FeK$\alpha$ line at 6.4~keV with an EW=0.22~keV does exist. They
suggest the presence of an accretion disk in addition to the jet for
explaining the origin of the X-ray emission. Chiaberge et al. (2003)
modeled the spectrum from $\gamma$-ray to radio
frequencies and found that it is consistent with a Synchrotron Self
Compton model with an unexpected high resemblance to blazar-like
objects. This model together with the dispute on the existence of
FeK$\alpha$ leads Evans et al. (2005) to favor the relativistic jet
emission as the main component of the observed emission.  Our data are
in a remarkable good agreement with the ones reported by Gliozzi et
al. (2003).
\smallskip

{\bf NGC\,6240 (IC\,4625, UGC\,10592, 4C\,02.44).}  
Komossa et al. (2003) discovered a binary AGN in the
galaxy coincident  with the optical nucleus. They appear as compact
unresolved at energies between 2.5-8~keV. The spectroscopic analysis
shows a very hard radiation  for both nuclei, with
$\Gamma$=0.2
for one to the South and 0.9
for one to the North-East one. In both nuclei the FeK$\alpha$ is
present. Satyapal et al. (2004) class it
as an object that reveal a hard nuclear point source embedded in soft
diffuse emission. Ptak et al. (2003) pointed out to the complexity of
the nuclear spectrum of this galaxy and construct a more complex model
which, in addition to the standard MEKAL and power-law component, they
also includes a gaussian fit for the FeK$\alpha$ and a Compton
reflection component with different column densities. To give an idea
of the complexity of the source let us point out that Boller et
al. (2003) best modeled the FeK$\alpha$ line as resolved into 3 narrow
lines: neutral FeK$\alpha$ at 6.4~keV, an ionized line at 6.7~keV and
a blend of higher ionized lines (FeXXVI and the Fe K$\beta$ line) at
7.0~keV. For consistence with the statistical analysis we have modeled
the continuum spectrum with a combination of thermal plus a power-law
component without taken into account the complex FeK$\alpha$ line.
High absorption is derived for this source from both the spectral
fitting and the estimation from color-color diagrams (Table 6).
\smallskip

{\bf IRAS\,17208-0014.}  
The X-ray nuclear emission of this Infrared Ultra-luminous Galaxy 
appears to be unresolved  at high energies (Fig. \ref{fig:clasif}). Rissaliti et al. (2000)
analyzed luminous IR galaxies in X-ray  with \emph{BeppoSAX} to investigate
the 2-10~keV nature of their emission and they classified  it as a star
forming galaxy with a quite large X-ray luminosity (L(2-10~keV)=1
$\times$ 10$^{42}$ erg~s$^{-1}$). Franceschini et al. (2003) report
\emph{XMM-Newton} data for a sample of 10 ULIRGs and found that for this
galaxy the observations are equally  consistent with a model of a
thermal plasma with a temperature kT=0.75~keV
plus a power-law
component with $\Gamma$=2.26
and N$\rm{_H}$=1.1$\times$
10$^{22}$cm$^{-2}$, and a thermal component with a temperature
kT=0.74~keV
plus a cut-off power-law component with $\Gamma$=1.30
and N$\rm{_H}$=2.6$\times$ 10$^{21}$cm$^{-2}$, leading in both cases
to a similar luminosity of the order of a few times 10$^{41}$erg
s$^{-1}$. They suggested, based on the lack of FeK$\alpha$ and the
close value between the SFR estimated through the Far IR emission and
the X-ray emission, that X-ray emission has an starburst origin. We
did not tried to fit the spectrum due to low counts.  From the
position in the color-color diagrams, this galaxy seems to be
consistent with high column density and a mix model with power-law
index between 1.6 and 2.0 and temperature in the range 0.6-0.8~keV.
Ptak et al. (2003) analyze the \emph{Chandra} data on this object and
obtain that the best fit to the global spectrum is provided by a
combined power-law ($\Gamma$=1.68) and thermal (kT=0.35 keV) with
N$\rm{_H}$=0.52$\times$ 10$^{22}$cm$^{-2}$model. The nuclear
luminosity is estimated to be L(2-10 keV)=4.2$\times$ 10$^{41}$
erg~s$^{-1}$, a factor of 3 brighter than the one we estimate (see
Table \ref{tab:lumflux}).
\smallskip

{\bf NGC\,6482 (UGC\,11009).}  This galaxy is the brightest member of
a fossil group. Khosroshashi et al. (2004) analyze the temperature
profile of the group but not for the individuals. The \emph{Chandra}
data on this source show no hard nuclear source
(Fig. \ref{fig:clasif}) associated with the compact radio source
detected by Goudfroij et al. (1994). Our spectral analysis show that
the data are consistent with a thermal plasma at kT=0.68~keV. This is
the only galaxy for which the nuclear spectrum is better fitted by a
single thermal component.
\smallskip


{\bf NGC\,7130 (IC\,5135).}  This galaxy shows a well defined nuclear
source at high X-ray energies (Fig. \ref{fig:clasif}).  Since most of
the UV emission is spectrally characteristic of star formation (Thuan
1984, Gonzalez-Delgado et al. 1998), Levenson et al. (2005) tried to
decompose the AGN and Starburst contributions and found that the AGN
contribution manifested mainly at higher energies, larger than 3~keV.
They found that the obscuration of the nucleus is Compton-thick what
prevents the detection of the intrinsic emission in the \emph{Chandra}
bandpass below 8~keV. We recall that our spectral fitting is not
statistically acceptable for this source.
\smallskip

{\bf NGC\,7331 (UGC\,12113).}  
Stockdale et al. (1998) and Roberts \& Warwicl (2000) 
pointed out, based in \emph{ROSAT}
data, to the AGN nature of this galaxy.  Nevertheless, the hard X-ray data from
\emph{Chandra} do not show any evidence of a nuclear source, being
very diffuse at high energies (Fig. \ref{fig:clasif}). 
Note that Filho et al. (2004) describe this galaxy as hosting a hard
(2-10 keV) X-ray nucleus, but Satyapal et al. (2004) 
class it as an object exhibiting
multiple, hard off-nuclear point sources of comparable brightness to
the nuclear source. Our estimated parameters are consistent with a
spectral index of 2-2.6 and temperature of 0.7~keV.  Recently Gallo et
al. (2006) present
\emph{XMM-Newton} data on the source and found that the spectrum is
consistent with a thermal component at kT=0.49~keV
plus a power-law with
$\Gamma$=1.79
giving a luminosity that it is a factor of 10 larger than our
estimation. The reasons for this difference are not clear. 
Nevertheless, the estimation of the luminosity by Satyapal et
al. (2004) for an intrinsic power slope of 1.8 is in perfect agreement
with ours, hinting to resolution effects being important to explain
the difference with the work by Gallo et al. (2006).
\smallskip

{\bf IC\,1459 (IC\,5265).}  This galaxy presents an unresolved nuclear
source on top of a diffuse halo at high energies
(Fig. \ref{fig:clasif}), in agreement with the classification by
Satyapal et al. (2004). It hosts a Super-massive black hole (2
$\times$ 10$^9$ M$\odot$, Cappellari et al. 2002) but with rather
moderate nuclear activity. Fabbiano et al. (2003) obtain that it shows
a rather weak (L(2-10~keV)=8.0$\times$ 10$^{40}$ erg~s$^{-1}$ )
unabsorbed nuclear X-ray source with $\Gamma$=1.88
and faint FeK$\alpha$ line at 6.4~keV. These
characteristics describe a normal AGN radiating at sub-Eddington
luminosities, at 3$\times$ 10$^{-7}$ below the Eddington
limit. They suggest that ADAF solutions can explain the X-ray spectrum
but these models failed to explain the high radio-power of its compact
source (Drinkwater et al. 1997). Our fitting parameters are in a
remarkably good agreement with theirs ($\Gamma$=1.89,
kT=0.30~keV
and L(2-10~keV)=3.6$\times$ 10$^{40}$ erg~s$^{-1}$).

\onecolumn
\newpage  
\begin{table}[!h]
\caption{Observational details.}\label{tab:obsdata}
\begin{tabular}{l c c c c c c c c c c@{~~~~~~}}  
\hline
\noalign{\smallskip}
\multicolumn {1}{l}{}
& \multicolumn {2}{c}{X-ray Position}
& \multicolumn {1}{l}{X-ray}
& \multicolumn {1}{c}{Offset}
& \multicolumn {3}{c}{X-ray Obs.}
& \multicolumn {3}{c}{Optical/HST Obs.~~~~~~~~} \\
 \multicolumn {1}{l}{Name~$^{(1)}$ }
& \multicolumn {1}{c}{$\rm{\alpha}~^{(2)}$}
& \multicolumn {1}{c}{$\rm{\delta}~^{(3)}$}
& \multicolumn {1}{c}{Radii~$^{(4)}$}
& \multicolumn {1}{c}{2MASS~$^{(5)}$}
& \multicolumn {1}{c}{Obs.ID~$^{(6)}$}
& \multicolumn {1}{c}{Expt.~$^{(7)}$}
& \multicolumn {1}{c}{Filter~$^{(8)}$}
& \multicolumn {1}{c}{Prop. ID~$^{(9)}$}
& \multicolumn {1}{c}{Expt.~$^{(10)}$}\\
& (2000)& (2000) & (arcsec)  & (arcsec) & & (ks) & & & (s) \\
\hline \\
 NGC\,0315\dotfill	 &00 57 48.9  &   +30 21 08.7& 1.968 &   0.81  &    4156    &  52.3    & F814W & 6673 & 460 \\
 ARP\,318A\dotfill	 &02 09 38.5  & $-$10 08 45.4& 4.920 &   1.26  &     923    &  12.5    & ...   & ...  & ...  \\  
 ARP\,318B\dotfill	 &02 09 42.7  & $-$10 11 02.2& 2.952 &   0.54  &     923    &  12.5    & ...   & ...  & ...  \\ 
 NGC\,1052\dotfill	 &02 41 04.8  & $-$08 15 20.6& 1.476 &   0.57  &     385    &  10.9    & F555W & 3639 & 1000 \\
 NGC\,2681\dotfill       &08 53 32.8  &   +51 18 48.8& 2.460 &   0.83  &    2060    &  78.3    & F555W & 4854 & 1000 \\  
 UGC\,05101\dotfill      &09 35 51.7  &   +61 21 11.1& 2.952 &   0.61  &    2033    &  48.2    & F814W & 6346 & 800 \\
 NGC\,3226\dotfill       &10 23 27.0  &   +19 53 54.6& 1.476 &   0.26  &     860    &  46.3    & F702W & 6357 & 1000 \\
 NGC\,3245\dotfill       &10 27 18.3  &   +28 30 27.2& 1.968 &   0.59  &    2926    &	9.6    & F702W & 7403 & 140 \\
 NGC\,3379\dotfill	 &10 47 49.6  &   +12 34 53.7& 1.968 &   0.20  &    1587    &  31.3    & F814W & 5512 & 1340 \\
 NGC\,3507\dotfill       &11 03 25.3  &   +18 08 07.3& 2.460 &   1.27  &    3149    &  38.0    & F606W & 5446 & 160 \\
 NGC\,3607\dotfill       &11 16 54.6  &   +18 03 04.4& 2.952 &   2.22  &    2073    &  38.4    & F814W & 5999 & 320 \\
 NGC\,3608\dotfill       &11 16 58.9  &   +18 08 53.7& 1.968 &   1.53  &    2073    &  38.4    & F814W & 5454 & 460 \\
 NGC\,3628\dotfill       &11 20 16.9  &   +13 35 22.8& 2.952 &   1.60  &    2039    &  54.8    & ...   & ...  & ...  \\
 NGC\,3690B\dotfill      &11 28 30.9  &   +58 33 40.5& 3.198 &   2.64  &    1641    &  24.3    & F814W & 8602 & 700 \\
 NGC\,4111\dotfill       &12 07 03.1  &   +43 03 57.3& 2.952 &   1.95  &    1578    &  14.8    & F702W & 6785 & 600 \\
 NGC\,4125\dotfill       &12 08 06.0  &   +65 10 27.6& 2.460 &   0.81  &    2071    &  61.9    & F814W & 6587 & 2100 \\
 NGC\,4261\dotfill       &12 19 23.2  &   +05 49 30.4& 3.444 &   1.49  &     834    &  31.3    & F702W & 5476 & 280 \\
 NGC\,4314\dotfill       &12 22 32.1  &   +29 53 44.8& 5.904 &   2.71  &    2062    &  14.5    & F814W & 6265 & 600 \\
 NGC\,4374\dotfill	 &12 25 03.7  &   +12 53 13.2& 1.476 &   0.90  &     803    &  28.2    & F814W & 6094 & 520 \\
 NGC\,4395\dotfill	 &12 25 48.9  &   +33 32 48.4& 1.476 &   0.78  &    5301    &  26.3    & F814W & 6464 & 2160 \\
 NGC\,4410A\dotfill      &12 26 28.2  &   +09 01 10.8& 1.968 &   1.00  &    2982    &  34.8    & F606W & 5479 & 500 \\  
 NGC\,4438\dotfill       &12 27 45.6  &   +13 00 33.0& 2.952 &   2.26  &    2883    &  24.6    & F814W & 6791 & 1050 \\
 NGC\,4457\dotfill       &12 28 59.0  &   +03 34 14.3& 2.952 &   0.45  &    3150    &  35.6    & ...   & ...  & ...  \\
 NGC\,4459\dotfill       &12 29 00.0  &   +13 58 41.7& 1.968 &   1.17  &    2927    &	9.8    & F814W & 5999 & 320 \\
 NGC\,4486\dotfill	 &12 30 49.4  &   +12 23 28.3& 2.460 &   0.35  &    2707    &  98.7    & F814W & 6775 & 1480 \\
 NGC\,4494\dotfill       &12 31 24.1  &   +25 46 29.8& 1.968 &   0.94  &    2079    &  15.8    & F814W & 6554 & 1800 \\
 NGC\,4552\dotfill	 &12 35 39.8  &   +12 33 23.3& 2.460 &   1.91  &    2072    &  54.3    & F814W & 6099 & 1500 \\
 NGC\,4579\dotfill	 &12 37 43.5  &   +11 49 05.5& 1.968 &   1.53  &     807    &  30.6    & F791W & 6436 & 600 \\
 NGC\,4594\dotfill	 &12 39 59.4  & $-$11 37 23.0& 1.476 &   0.71  &    1586    &  18.5    & F814W & 5512 & 1470 \\
 NGC\,4596\dotfill       &12 39 56.0  &   +10 10 35.3& 3.690 &   1.61  &    2928    &	9.2    & F606W & 5446 & 160 \\
 NGC\,4636\dotfill       &12 42 49.8  &   +02 41 15.9& 2.214 &    ...  &    4415    &  74.1    & F814W & 8686 & 400 \\  
 NGC\,4676A\dotfill      &13 30 03.3  &   +47 12 41.4& 3.936 &   1.31  &    2043    &  27.9    & F814W & 8669 & 320 \\
 NGC\,4676B\dotfill      &13 30 04.6  &   +47 12 08.9& 3.936 &    ...  &    2043    &  27.9    & F814W & 8669 & 320 \\
 NGC\,4696\dotfill       &12 48 48.8  & $-$41 18 43.3& 3.936 &   7.51  &    1560    &  47.7    & F814W & 8683 & 1000 \\
 NGC\,4698\dotfill	 &12 48 22.9  &   +08 29 14.6& 2.214 &   0.71  &    3008    &  29.4    & F814W & 9042 & 460 \\
 NGC\,4736\dotfill	 &12 50 53.1  &   +41 07 13.2& 1.476 &   0.82  &     808    &  46.4    & F814W & 9042 & 460 \\  
 NGC\,5055\dotfill	 &13 15 49.2  &   +42 01 45.9& 1.476 &   1.33  &    2197    &  27.7    & F814W & 9042 & 460 \\
 NGC\,5194\dotfill	 &13 29 52.8  &   +47 11 40.4& 3.444 &   2.87  &    3932    &  45.2    & ...   & ...  & ...  \\
 MRK\,0266NE\dotfill     &13 38 17.9  &   +48 16 41.1& 1.968 &   0.84  &    2044    &  17.4    & F606W & 5479 & 500 \\
 UGC\,08696\dotfill      &13 44 42.1  &   +55 53 13.1& 1.968 &   0.66  &     809    &  40.4    & F814W & 8645 & 800 \\
 CGCG\,162-010\dotfill   &13 48 52.5  &   +26 35 36.3& 3.936 &   1.59  &     493    &  19.6    & F702W & 5212 & 300 \\
 NGC\,5746\dotfill       &14 44 56.0  &   +01 57 18.1& 1.476 &   1.06  &    3929    &  36.8    & F814W & 9046 & 800 \\
 NGC\,5846\dotfill       &15 06 29.2  &   +01 36 19.6& 4.920 &   1.13  &    4009    &  24.0    & F814W & 5920 & 2300 \\
 NGC\,5866\dotfill	 &15 06 29.5  &   +55 45 46.1& 4.428 &   2.22  &    2879    &  31.9    & ...   & ...  & ...  \\
 NGC\,6251\dotfill       &16 32 31.9  &   +82 32 15.7& 1.968 &   1.68  &     847    &  25.4    & F814W & 6653 & 1000 \\
 NGC\,6240\dotfill       &16 52 58.9  &   +02 24 02.6& 1.680 &   0.99  &    1590    &  36.7    & F814W & 6430 & 1200 \\
 IRAS\,17208-0014\dotfill &17 23 22.0 &   +00 17 00.6& 2.460 &   1.04  &    2035    &  48.4    & F814W & 6346 & 800 \\
 NGC\,6482\dotfill       &17 51 48.8  &   +23 04 18.9& 2.460 &   0.46  &    3218    &  19.1    & ...   & ...  & ...  \\
 NGC\,7130\dotfill       &21 48 19.5  & $-$34 57 05.0& 1.968 &   0.73  &    2188    &  36.9    & F606W & 5479 & 1900 \\
 NGC\,7331\dotfill       &22 37 04.0  &   +34 24 56.0& 1.968 &   1.75  &    2198    &  29.5    & F814W & 7450 & 170 \\
 IC\,1459\dotfill        &22 57 10.6  & $-$36 27 43.6& 2.952 &   1.81  &    2196    &  51.8    & F814W & 5454 & 460 \\ 
\hline													     
\end{tabular}
\end{table}
\topmargin=-30pt
\begin{table}[!h]
\begin{center}
\caption{Host galaxy properties.}\label{tab:catdata}
\begin{tabular}{@{~}l@{~~~~~}c@{~~~~~}c@{~~~~~}c@{~~~~~}c@{~~~~~}c@{~~~~~}c@{~~~~~}c@{~~~~~}l@{~~~~~}} \hline \hline \\ 
Name~$^{(1)}$  & z~$^{(2)}$ & Dist.~$^{(3)}$ & Spatial Scale~$^{(4)}$ & Source radii~$^{(5)}$ &{\it B~$^{(6)}$ } & E(B-V)~$^{(7)}$ & Morph. Type~$^{(8)}$  \\ 
	&  &	    &	(pc/arcsec)&	(pc)&	&	& \\ \hline \\ 
 NGC\,0315\dotfill	   	  & 0.016465 &  65.8$^{(b)}$ & 319.0 &   627.8    &12.2  & 0.065 & E		    \\     
 ARP\,318A\dotfill	   	  & 0.013586 &  54.3$^{(b)}$ & 263.3 &  1295.4    &12.91 & 0.025 & SAB(r)ab;pec    \\ 
 ARP\,318B\dotfill	   	  & 0.012889 &  54.3$^{(b)}$ & 263.3 &   777.3    &13.69 & 0.025 & (R')Sa;pec  \\
 NGC\,1052\dotfill	   	  & 0.004903 &  19.4$^{(c)}$ &  94.1 &   138.9    &11.41 & 0.027 & E4	   \\ 
 NGC\,2681\dotfill         	  & 0.002308 &  17.2$^{(c)}$ &  83.4 &   205.2    &11.09 & 0.023 & (R')SAB(rs)0/a \\
 UGC\,05101\dotfill        	  & 0.039390 & 157.3$^{(b)}$ & 762.6 &  2251.2    &15.2  & 0.033 & S?	   \\ 
 NGC\,3226\dotfill         	  & 0.003839 &  23.6$^{(c)}$ & 114.4 &   168.9    &12.3  & 0.023 & E2;pec	   \\
 NGC\,3245\dotfill         	  & 0.004530 &  20.9$^{(c)}$ & 101.3 &   199.4    &11.7  & 0.025 & SA(r)0	    \\
 NGC\,3379\dotfill	   	  & 0.003069 &  10.6$^{(c)}$ &  51.4 &   101.2    &10.24 & 0.024 & E1	    \\
 NGC\,3507\dotfill         	  & 0.003266 &  19.8$^{(d)}$ &  96.0 &   236.2    &11.73 & 0.024 & SB(s)b	    \\
 NGC\,3607\dotfill         	  & 0.003119 &  22.8$^{(c)}$ & 110.5 &   326.2    &10.82 & 0.021 & SA(s)0	    \\
 NGC\,3608\dotfill         	  & 0.003696 &  22.9$^{(c)}$ & 111.0 &   218.4    &11.7  & 0.021 & E2	    \\
 NGC\,3628\dotfill         	  & 0.002812 &   7.7$^{(d)}$ &  37.3 &   110.1    &14.8  & 0.027 & Sb;pec;sp	 \\
 NGC\,3690B\dotfill        	  & 0.010411 &  41.6$^{(b)}$ & 201.7 &   645.0    &...   & 0.017 & GPair    \\
 NGC\,4111\dotfill         	  & 0.002692 &  15.0$^{(c)}$ &  72.7 &   214.6    &11.63 & 0.015 & SA(r)0+:;sp   \\
 NGC\,4125\dotfill         	  & 0.004523 &  23.9$^{(c)}$ & 115.9 &   285.1    &10.65 & 0.019 & E6;pec	    \\
 NGC\,4261\dotfill         	  & 0.007372 &  29.4$^{(c)}$ & 142.5 &   490.8    &11.41 & 0.018 & E2-3     \\
 NGC\,4314\dotfill         	  & 0.003212 &   9.7$^{(d)}$ &  47.0 &   277.5    &11.43 & 0.025 & SB(rs)a	   \\
 NGC\,4374\dotfill	   	  & 0.003336 &  18.4$^{(c)}$ &  89.2 &   131.7    &12.1  & 0.040 & I	   \\
 NGC\,4395\dotfill	   	  & 0.001064 &   4.2$^{(e)}$ &  20.4 &    30.1    &10.64 & 0.017 & SA(s)m	   \\	     
 NGC\,4410A\dotfill        	  & 0.025174 & 100.5$^{(b)}$ & 487.2 &   958.8    &14.92 & 0.024 & S0;pec	     \\
 NGC\,4438\dotfill         	  & 0.000237 &  16.8$^{(d)}$ &  81.4 &   240.3    &11.02 & 0.028 & SA(s)0/a;p:   \\
 NGC\,4457\dotfill         	  & 0.002942 &  17.4$^{(d)}$ &  84.4 &   249.1    &11.76 & 0.022 & (R)SAB(s)0/a  \\
 NGC\,4459\dotfill         	  & 0.004036 &  16.1$^{(c)}$ &  78.1 &   153.7    &11.32 & 0.046 & SA(r)0+	     \\
 NGC\,4486\dotfill	   	  & 0.004276 &  16.1$^{(c)}$ &  78.1 &   192.1    &9.59  & 0.022 & E+0-1;pec	 \\
 NGC\,4494\dotfill         	  & 0.004416 &  17.1$^{(c)}$ &  82.9 &   163.1    &10.71 & 0.021 & E1-2      \\
 NGC\,4552\dotfill	   	  & 0.001071 &  15.4$^{(c)}$ &  74.7 &   183.8    &10.73 & 0.041 & E		     \\
 NGC\,4579\dotfill	   	  & 0.005067 &  16.8$^{(d)}$ &  81.4 &   160.2    &10.48 & 0.041 & SAB(rs)b	 \\
 NGC\,4594\dotfill	   	  & 0.003639 &   9.8$^{(c)}$ &  47.5 &    70.1    &8.98  & 0.051 & SA(s)a      \\
 NGC\,4596\dotfill         	  & 0.006251 &  16.8$^{(d)}$ &  81.4 &   300.4    &11.35 & 0.022 & SB(r)0+     \\
 NGC\,4636\dotfill         	  & 0.003653 &  14.7$^{(c)}$ &  71.3 &   157.9    &10.43 & 0.028 & E/S0\_1     \\
 NGC\,4676A\dotfill        	  & 0.022059 &  88.0$^{(b)}$ & 426.6 &  1679.1    &14.7  & 0.017 & Irr       \\
 NGC\,4676B\dotfill        	  & 0.022039 &  88.0$^{(b)}$ & 426.6 &  1679.1    &14.4  & 0.017 & SB(s)0/a;pec  \\
 NGC\,4696\dotfill         	  & 0.009867 &  35.5$^{(c)}$ & 172.1 &   677.4    &11.39 & 0.113 & E+1;pec     \\
 NGC\,4698\dotfill	   	  & 0.003342 &  16.8$^{(d)}$ &  81.4 &   180.2    &1.46  & 0.026 & SA(s)ab	      \\	
 NGC\,4736\dotfill	   	  & 0.001027 &   5.2$^{(c)}$ &  25.2 &    37.2    &8.99  & 0.018 & (R)SA(r)ab  \\
 NGC\,5055\dotfill	   	  & 0.001681 &   7.2$^{(d)}$ &  34.9 &    51.5    &9.31  & 0.018 & SA(rs)bc	     \\
 NGC\,5194\dotfill	   	  & 0.001544 &   7.7$^{(c)}$ &  37.3 &   128.5    &8.96  & 0.035 & SA(s)bc;pec    \\
 MRK\,0266NE\dotfill       	  & 0.028053 & 112.0$^{(b)}$ & 543.0 &  1068.6    &14.1  & 0.013 & Compact;pec:   \\
 UGC\,08696\dotfill        	  & 0.037780 & 150.9$^{(b)}$ & 731.6 &  1439.8    &15.07 & 0.008 & Ring galaxy    \\	    
 CGCG\,162-010.~.~.~.~.~.\dotfill & 0.063260 & 252.6$^{(b)}$ &1224.6 &  4820.0    &15.2  & 0.013 & cD;S0?	    \\
 NGC\,5746\dotfill       	  & 0.005751 &  29.4$^{(d)}$ & 142.5 &   210.3    &11.29 & 0.040 & SAB(rs)b?;sp   \\
 NGC\,5846\dotfill  		  & 0.006078 &  24.9$^{(c)}$ & 120.7 &   593.8    &11.05 & 0.055 & E0-1     \\
 NGC\,5866\dotfill		  & 0.002242 &  15.4$^{(c)}$ &  74.7 &   330.8    &10.74 & 0.013 & S0\_3    \\
 NGC\,6251\dotfill      	  & 0.023016 &  91.9$^{(b)}$ & 445.5 &   876.7    &13.64 & 0.087 & E		    \\
 NGC\,6240\dotfill                & 0.024480 &  97.8$^{(b)}$ & 474.1 &   796.5    &13.8  & 0.076 & I0:;pec	    \\
 IRAS\,17208-0014\dotfill         & 0.042810 & 171.0$^{(b)}$ & 829.0 &  2039.3    &15.1  & 0.344 & ...      \\
 NGC\,6482\dotfill                & 0.013176 &  52.6$^{(b)}$ & 255.0 &   627.3    &12.35 & 0.099 & E		    \\
 NGC\,7130\dotfill                & 0.016151 &  64.5$^{(b)}$ & 312.7 &   615.4    &12.98 & 0.029 & Sa;pec	   \\
 NGC\,7331\dotfill                & 0.002739 &  15.1$^{(a)}$ &  73.2 &   144.1    &10.35 & 0.091 & SA(s)b	\\
 IC\,1459\dotfill                 & 0.005641 &  22.0$^{(d)}$ & 106.7 &   315.0    &10.97 & 0.016  & E3      \\ \hline
\end{tabular}								  
\end{center}								  
{\sc notes.}--Properties are extracted from the Carrillo et al. 1999 and references therein.  
(a) Ferrarese et al. 2000;
(b) from cosmology assuming $\rm{H_{o}}=75 Km~Mpc^{-1}~s^{-1}$;
(c) Tonry et al. 2001;
(d) Tully R.B. 1998, Nearby Galaxies Catalog (Cambridge: Cambridge Univ. Press); and 
(e) Karachentsev I.D. \& Drodovsky I.O. 1998.

\end{table}
\onecolumn
\begin{table}
\caption{Results from spectral fitting of the five models tested.}\label{tab:fittings_anex}
\twocolumn
  \begin{tabular}{l@{}c@{~}c@{~}c@{~}c@{~}c@{~}} \hline \hline \\ 
Name~$^{(1)}$	  &  Model~$^{(2)}$ & $\rm{N_H}$~$^{(3)}$ &  $\rm{\Gamma}$~$^{(4)}$ & kT~$^{(5)}$   & $\rm{\chi^{2}}$/d.o.f~$^{(6)}$ \\
          &	   & $(\rm{10^{22}cm^{-2}})$  &                &(keV) &                      \\  \hline \\
NGC\,0315\dotfill	  & RS     &	  0.23&... & 63.99& 	285.15/163   \\
			  & ME     &	  0.22&... & 79.89& 	284.03/163   \\
			  & PL     &	  0.20&1.18&...   & 	282.77/163  \\
 			  &RS+PL(*)&	  0.71&1.48&  0.22& 	188.77/161  \\
			  &ME+PL   &	  0.48&1.30&  0.27& 	213.89/161  \\ \hline
NGC\,2681\dotfill	  & RS     &	  0.51&... &  0.64& 	 78.02/28   \\
			  & ME     &	  0.45&... &  0.58& 	 77.75/28   \\
			  & PL     &	  0.28&3.68&  ... & 	 82.58/28   \\
			  &RS+PL   &	  0.04&1.68&  0.76& 	  9.40/26   \\
			  &ME+PL(*)&	  0.08&1.74&  0.66& 	 13.94/26   \\ \hline
NGC\,3690B\dotfill	  & RS     &	  0.79&... &  1.40& 	118.70/25   \\
			  & ME     &	  1.09&... &  1.10& 	112.04/25   \\
			  & PL(*)  &	  0.02&1.45&  ... & 	 21.81/25   \\
			  &RS+PL   &	  0.10&1.45&  0.98& 	 20.28/23   \\
 			  &ME+PL   &	  0.07&1.43&  1.08& 	 20.13/23   \\ \hline
 NGC\,4374\dotfill	  & RS     &	  0.97&... &  0.64& 	168.94/33   \\
			  & ME     &	  0.93&... &  0.62& 	179.05/33   \\
			  &PL(*)   &	  0.14&2.07&  ... & 	 32.20/33   \\
			  &RS+PL   &	  0.12&1.80&  0.77& 	 18.76/31   \\
			  &ME+PL   &	  0.14&1.83&  0.64& 	 18.34/31   \\ \hline
 NGC\,4395\dotfill	  & RS     &	  2.73&... & 28.01& 	120.52/106  \\
			  & ME     &	  2.77&... & 24.01& 	119.41/106  \\
			  &PL(*)   &	  2.87&1.44&  ... & 	123.68/106  \\
			  &RS+PL   &	  2.93&1.48&  0.03& 	123.62/104  \\
			  &ME+PL   &	  2.97&1.51&  0.08& 	123.69/104  \\ \hline
NGC\,4410A\dotfill	  & RS     &	  0.92&... &  0.60& 	337.45/45   \\
			  & ME     &	  0.89&... &  0.58& 	353.98/45   \\
			  & PL(*)  &	  0.01&1.75&  ... & 	 33.11/45   \\
			  &RS+PL   &	  0.01&1.69&  0.58& 	 31.21/43   \\
			  &ME+PL   &	  0.01&1.68&  0.50& 	 30.32/43   \\ \hline
 NGC\,4438\dotfill	  & RS     &	  0.43&... &  0.66& 	 68.13/36   \\
			  & ME     &	  0.41&... &  0.57& 	 53.97/36   \\
			  & PL     &	  0.89&7.69&   ...& 	148.92/36   \\
			  &RS+PL(*)&	  0.11&2.13&  0.78& 	 32.27/34   \\  				   
			  &ME+PL   &	  0.23&2.42&  0.60& 	 29.04/34   \\ \hline				   
 NGC\,4457\dotfill	  & RS     &	  0.51&... &  0.63& 	 74.99/20   \\
			  & ME     &	  0.43&... &  0.52& 	 78.02/20   \\
			  & PL     &	  0.32&4.08&  ... & 	 68.80/20   \\
			  &RS+PL(*)&	  0.07&1.98&  0.66& 	 18.68/18   \\
			  &ME+PL   &	  0.13&1.96&  0.53& 	 19.23/18   \\ \hline
 NGC\,4494\dotfill	  & RS     &	  0.78&... &  0.61& 	 27.90/7    \\
			  & ME     &	  0.73&... &  0.56& 	 29.40/7    \\
			  & PL     &	  0.12&2.38&  ... & 	 16.47/7    \\
			  &RS+PL   &	  0.01&1.13&  0.82& 	  5.03/5    \\
			  &ME+PL(*)&	  0.11&1.37&  0.63& 	  5.59/5    \\ \hline
 NGC\,4552\dotfill	  & RS     &	  0.75&... &  0.66& 	354.09/63   \\
			  & ME     &	  0.71&... &  0.61& 	368.32/63   \\
			  & PL     &	  0.12&2.39&  ... & 	117.28/63   \\
			  &RS+PL(*)&	  0.04&1.81&  0.83& 	 57.19/61   \\
			  &ME+PL   &	  0.05&1.82&  0.79& 	 70.58/61   \\ \hline
 NGC\,4594\dotfill~.~.~   &RS &      0.19&... & 18.77& 	115.89/87   \\
			  & ME     &	  0.19&... & 18.49& 	116.02/87   \\
			  &PL(*)   &	  0.20&1.41&  ... & 	118.72/87   \\
			  &RS+PL   &	  0.22&1.41&  0.45& 	117.65/85   \\
			  &ME+PL   &	  0.25&1.43&  0.39& 	117.03/85   \\ \hline
 NGC\,4696\dotfill	  & RS     &	  0.37&... &  0.70&    26.71/17   \\
			  & ME     &	  0.35&... &  0.62&    24.47/17   \\
			  &PL(*)   &	  0.13&4.26&  ... &    21.98/17   \\
			  &RS+PL   &	  0.41&4.67&  0.27&    19.94/15   \\
			  &ME+PL   &	  0.62&2.33&  0.18&	5.80/15   \\ \hline
\end{tabular}
\end{table}
\begin{table}
~~\newline
~~
\newline
~~
\newline
~~
\newline
  \begin{tabular}{l@{}c@{~}c@{~}c@{~}c@{~}c@{~}} \hline \hline \\ 
Name~$^{(1)}$	  &  Model~$^{(2)}$ & $\rm{N_H}$~$^{(3)}$ &  $\rm{\Gamma}$~$^{(4)}$ & kT~$^{(5)}$   & $\rm{\chi^{2}}$/d.o.f~$^{(6)}$ \\
          &	   & $(\rm{10^{22}cm^{-2}})$  &                &(keV) &                      \\  \hline \\
 NGC\,4736\dotfill	  & RS     &	  0.82&... &  0.59&   635.49/81   \\
			  & ME     &	  0.71&... &  0.56&   673.96/81   \\
			  & PL     &	  0.12&2.38& ...  &   130.19/81   \\
			  &RS+PL   &	  0.08&2.02&  0.66&    71.96/79   \\
			  &ME+PL(*)&	  0.08&2.00&  0.60&    72.90/79   \\ \hline
 NGC\,5194\dotfill        & RS     &	  0.09&... &  0.68&   185.59/50   \\
			  & ME     &	  0.09&... &  0.60&   114.83/50   \\
			  & PL     &	  0.82&8.00&  ... &   340.52/50   \\
			  &RS+PL   &	  0.02&2.98&  0.69&    68.32/48   \\
			  &ME+PL(*)&	  0.03&2.67&  0.61&    49.74/48   \\ \hline
 UGC\,08696\dotfill	  & RS     &	  0.09& ...&  8.72&    88.79/42   \\
			  & ME     &	  0.09& ...&  9.51&    78.50/42   \\
			  & PL     &	  0.27&2.34&   ...&    72.20/42   \\
			  &RS+PL(*)&	  0.39&1.74&  0.75&    55.63/40   \\
			  &ME+PL   &	  0.38&1.70&  0.68&    56.81/40   \\ \hline
 CGCG\,162-010\dotfill    & RS    &	  0.51&... &  0.86&   117.41/63   \\
			  & ME     &	  0.90&... &  0.35&   118.54/63   \\
			  & PL     &	  0.39&3.99&  ... &    83.83/63   \\
			  &RS+PL(*)&	  0.09&2.44&  1.12&    59.95/61   \\
			  &ME+PL   &	  0.12&2.71&  1.33&    58.43/61   \\ \hline
 NGC\,5746\dotfill	  & RS     &	  0.64& ...& 63.97&    15.91/17   \\			 
			  & ME     &	  0.62& ...& 79.90&    15.90/17   \\			 
			  & PL(*)  &	  0.60&1.22&  ... &    15.66/17   \\
			  &RS+PL   &	  1.13&1.51&  0.11&    14.81/15   \\
			  &ME+PL   &	  0.95&1.43&  0.08&    15.79/15   \\ \hline
 NGC\,5846\dotfill	  & RS     &	  0.16&... &  0.65&    80.52/27   \\
			  & ME     &	  0.25&... &  0.48&    45.59/27   \\
			  & PL     &	  0.79&7.20&  ... &   160.99/27   \\
			  &RS+PL   &	  0.10&2.56&  0.63&    38.16/25   \\
			  &ME+PL(*)&	  0.12&1.65&  0.50&    21.91/25   \\ \hline
 NGC\,6251\dotfill	  & RS     &	  0.03&... & 32.62&    74.98/64   \\
			  & ME     &	  0.03&... & 31.18&    75.08/64   \\
			  & PL     &	  0.05&1.36&  ... &    73.08/64   \\
			  &RS+PL   &	  0.15&1.42& 0.49 &    67.85/62   \\
			  &ME+PL(*)&	  0.44&1.60& 0.26 &    65.27/62   \\  \hline
 NGC\,6240\dotfill	  & RS     &	  0.29&... & 48.10&   181.56/114  \\  
			  & ME     &	  0.30&... & 37.87&   181.24/114  \\
			  & PL     &	  0.31&1.31&  ... &   181.95/114  \\
			  &RS+PL(*)&	  1.11&1.03& 0.76 &   126.79/112  \\
			  &ME+PL   &	  1.17&1.08& 0.75 &   133.10/112  \\ \hline
 NGC\,6482\dotfill	  & RS     &	  0.30&... &  0.78&    15.90/18   \\
			  &ME(*)   &	  0.31&... &  0.68&    16.41/18   \\
			  & PL     &	  1.02&9.24&  ... &    41.40/18   \\
			  &RS+PL   &	  0.02&2.42&  0.87&    13.21/16   \\
			  &ME+PL   &	  0.02&2.64&  0.83&    14.51/16   \\ \hline
 NGC\,7130\dotfill	  & RS     &	  0.57&... &  0.63&   284.55/59   \\
			  & ME     &	  0.54&... &  0.51&   266.72/59   \\
			  & PL     &	  0.32&4.23&  ... &   192.78/59   \\
			  &RS+PL(*)&	  0.06&2.51&  0.82&    81.43/57   \\
			  &ME+PL   &	  0.12&2.74&  0.67&    90.75/57   \\ \hline
 IC\,1459\dotfill 	  & RS     &	  0.06& ...&  7.04&   341.31/168  \\
			  & ME     &	  0.06& ...&  7.16&   335.79/168  \\
			  & PL     &	  0.16&1.89&   ...&   290.02/168  \\
			  &RS+PL   &	  0.18&1.80&  0.61&   193.27/166  \\
			  &ME+PL(*)&	  0.28&1.89&  0.30&   189.47/166  \\ \hline
\end{tabular}
\end{table}

\onecolumn

\begin{table}[!h]
\caption{Result of the model fitting to the spectra of the SF subsample. The top line in each parameter corresponds to the whole
sample of 24 objects, bottom-left shows the result in the AGN-like
nuclei (19 objects) and bottom-right shows the result in the SB-like
objects (5 objects).}\label{tab:fits}
\begin{center}
  \begin{tabular}{lccc} \hline \hline \\ 
     & \texttt{Mean~$^{(1)}$} & \texttt{Median~$^{(2)}$} &   \texttt{Mean Stand Dev.~$^{(3)}$} \\ 
        	        &  AGN/SB    &  AGN/SB    &  AGN/SB   \\ \hline \\
Log($\rm{L_{X}}$) 	&  40.16     &   39.78	  &   0.96    \\
         	        &40.22/39.93 &40.07/39.54 &1.03/0.64  \\
        	        &            &      	  &           \\
$\rm{N_{H}}$	        &  0.35      &  0.12	  &   0.35    \\
$\rm{(10^{22}~cm^{-2})}$&0.40/0.15   &0.14/0.12   &0.41/0.06  \\
        	        &            &      	  &           \\
kT       	        &  0.64      &  0.66	  &   0.17    \\
(keV)     	 	&0.59/0.77   &0.66/0.78   &0.17/0.18  \\
        	        &            &      	  &           \\
$\rm{\Gamma}$           & 1.89       &  1.75      &   0.45    \\
     	 	        & 1.73/2.62  &1.74/2.44   &0.32/0.82  \\
        	        &            &      	  &           \\
\hline
\end{tabular}
\end{center}
\end{table}

\begin{table*}[!h]
\begin{center}
\caption{X-ray and HST results.}\label{tab:lumflux}
  \begin{tabular}{@{~~~~}l@{~~~~}c@{~~~~}c@{~~~~}c@{~~~~}c@{~~~~}c@{~~~~}c@{~~~~}} \hline \hline \\ 
Name~$^{(1)}$	  & Flux~$^{(2)}$    & Log($\rm{L_X}$(2.0--10.0) keV)~$^{(3)}$ 	&  Ref.~$^{(4)}$ & Obscuration~$^{(5)}$ & Classif.~$^{(6)}$  & HST-Classif.~$^{(7)}$  \\
		  &	$\rm{erg~cm^{-2}~s^{-1}}$ & $\rm{erg~s^{-1}}$		& 	    &  &  & \\  \hline \\
 NGC\,0315\dotfill	  	       &    -12.08 $_{    -0.06}^{	+ 0.04}$ &    41.64 $_{     -0.06}^{   +   0.03}$&     (f)   &SF+CD	     &      AGN      &    C	 \\
 ARP\,318A\dotfill	  	       &    -13.24& 39.48	&     (e)   &		    &	   SB	    &	       \\
 ARP\,318B\dotfill    	  	       &    -13.32& 40.20	&     (e)   &CD 	    &	   SB	    &	       \\
 NGC\,1052\dotfill	  	       &    -11.95& 40.78	&     (e)   &CD 	    &	   AGN      &	 C	\\
 NGC\,2681\dotfill	  	       &    -13.61 $_{    -0.20}^{     + 0.09}$ &	  38.94 $_{     -0.22}^{   +   0.10}$	&     (f)   &		    &	   AGN      &	 C	\\
 UGC\,05101\dotfill	  	       &    -13.32& 41.24	&     (e)   &		    &	   AGN      &	 C	\\
 NGC\,3226\dotfill	  	       &    -12.83& 39.62	&     (e)   &CD 	    &	   AGN      &	 C	\\
 NGC\,3245\dotfill	  	       &    -13.48& 39.08	&     (e)   &		    &	   SB	    &	 C     \\
 NGC\,3379\dotfill    	  	       &    -14.36& 37.89 	&     (e)   &    	    &	   SB       &	 C	\\
 NGC\,3507\dotfill	  	       &    -14.37& 38.31	&     (e)   &		    &	   SB	    &	 D     \\
 NGC\,3607\dotfill	  	       &    -14.05& 38.45	&     (e)   &CD 	    &	   SB	    &	 D     \\
 NGC\,3608\dotfill	  	       &    -13.97& 38.64	&     (e)   &CD 	    &	   SB	    &	 U     \\
 NGC\,3628\dotfill	  	       &    -13.61& 38.24	&     (e)   &CD 	    &	   AGN      &	 C	\\
 NGC\,3690B\dotfill	  	       &    -12.64 $_{    -0.06}^{     + 0.07}$ &	  40.63 $_{     -0.05}^{   +   0.07}$	&     (f)   &CD 	    &	   AGN      &	 C	\\
 NGC\,4111\dotfill	  	       &    -13.21& 39.33	&     (e)   &		    &	   AGN      &	 C	\\
 NGC\,4125\dotfill	  	       &    -13.91& 38.93	&     (e)   &		    &	   AGN      &	 C	\\
 NGC\,4261\dotfill	  	       &    -12.58& 40.65	&     (e)   &		    &	   AGN      &	 C	\\   
 NGC\,4314\dotfill	  	       &    -13.92& 37.94	&     (e)   &		    &	   SB	    &	 C     \\
 NGC\,4374\dotfill	  	       &    -13.03 $_{    -0.12}^{     + 0.06}$ &	  39.58 $_{     -0.10}^{   +   0.07}$	&     (f)   &SF 	    &	   AGN      &	 C	\\
 NGC\,4395\dotfill	  	       &    -11.61 $_{    -0.25}^{     + 0.06}$ &	  39.71 $_{     -0.26}^{   +   0.05}$		&     (f)   &SF 	    &	   AGN      &	 C	\\
 NGC\,4410A\dotfill	  	       &    -12.85 $_{    -0.06}^{     + 0.06}$ &    41.26 $_{	-0.06}^{   +   0.06}$	&     (f)   &		    &	   AGN      &	 C	\\
 NGC\,4438\dotfill	  	       &    -13.32 $_{    -0.37}^{     + 0.11}$ &    39.22 $_{	-0.28}^{   +   0.16}$	&     (f)   &SF 	    &	   SB	    &	 D     \\
 NGC\,4457\dotfill	  	       &    -13.51 $_{    -0.24}^{     + 0.14}$ &    39.05 $_{	-0.25}^{   +   0.13}$	&     (f)   &		    &	   AGN      &		\\
 NGC\,4459\dotfill	  	       &    -13.66& 38.83	&     (e)   &CD 	    &	   SB	    &	 C     \\
 NGC\,4486\dotfill	  	       &    -11.75& 40.75	&     (e)   &		    &	   AGN      &	 C	\\
 NGC\,4494\dotfill	  	       &    -12.93 $_{    -0.45}^{     + 0.19}$ &    39.62 $_{	-0.43}^{   +   0.17}$	&     (f)   &SF 	    &	   AGN      &	 C	\\
 NGC\,4552\dotfill	  	       &    -13.04 $_{    -0.09}^{	    + 0.06}$ &    39.41 $_{	-0.09}^{   +   0.06}$	&     (f)   &		    &	   AGN      &	 C	\\
 NGC\,4579\dotfill	  	       &    -11.38& 41.15	&     (e)   &		    &	   AGN      &	 C	\\
 NGC\,4594\dotfill        	       &    -11.99 $_{    -0.05}^{	    + 0.04}$ &    40.07 $_{	-0.06}^{   +   0.04}$	&     (f)   &SF 	    &	   AGN      &	 C	\\
 NGC\,4596\dotfill	  	       &    -13.88& 38.65	&     (e)   &		    &	   SB	    &	 C     \\
 NGC\,4636\dotfill	  	       &    -13.16& 39.25	&     (e)   &		    &	   SB	    &	 U     \\
 NGC\,4676A\dotfill	  	       &    -13.93& 40.06	&     (e)   &CD 	    &	   SB	    &	 D     \\
 NGC\,4676B\dotfill	  	       &    -13.78& 40.21	&     (e)   &CD 	    &	   SB	    &	 D     \\
 NGC\,4696\dotfill	  	       &    -13.42 $_{    -0.23}^{     + 0.13}$ &    39.78 $_{	-0.27}^{   +   0.09}$	&     (f)   &SF 	    &	   SB	    &	 C     \\
 NGC\,4698\dotfill	  	       &    -13.84& 38.72	&     (e)   &CD 	    &	   SB	    &	 D     \\
 NGC\,4736\dotfill	  	       &    -12.86 $_{    -0.07}^{     + 0.04}$ &    38.65 $_{	-0.08}^{   +   0.05}$	&     (f)   &		    &	   AGN      &	 C	\\
 NGC\,5055\dotfill	  	       &    -13.42& 38.37	&     (e)   &		    &	   AGN      &	 C	\\
 NGC\,5194\dotfill        	       &    -13.71 $_{    -0.19}^{     + 0.15}$ &    38.14 $_{	-0.21}^{   +   0.13}$	&     (f)   &		    &	   AGN      &		\\
 MRK\,0266NE\dotfill	 	       &    -13.22& 40.98	&     (e)   &		    &	   AGN      &	 C	\\ 
 UGC\,08696\dotfill		       &    -12.28 $_{    -0.21}^{     + 0.15}$ &    42.18 $_{	-0.22}^{   +   0.14}$	&     (f)   &SF 	    &	   AGN      &	 C	 \\
 CGCG\,162-010~.~.~.~.\dotfill         &    -13.23 $_{    -0.29}^{     + 0.19}$ &    41.73 $_{	-0.27}^{   +   0.23}$	&     (f)   &CD 	    &	   SB	    &	 D	\\
 NGC\,5746\dotfill	  	       &    -12.73 $_{    -0.56}^{     + 0.11}$ &    40.07 $_{	-0.54}^{   +   0.10}$	&     (f)   &CD 	    &	   AGN      &	 C	 \\
 NGC\,5846\dotfill	  	       &    -13.34 $_{    -0.59}^{     + 0.35}$ &    39.54 $_{	-0.59}^{   +   0.38}$	&     (f)   &SF 	    &	   SB	    &	 D	\\
 NGC\,5866\dotfill        	       &    -13.85& 38.80	&     (e)   &		    &	   SB	    &		\\
 NGC\,6251\dotfill	 	       &    -12.18 $_{    -0.04}^{     + 0.07}$ &    41.84 $_{	-0.05}^{   +   0.07}$	&     (f)   &SF 	    &	   AGN      &	 C	 \\
 NGC\,6240\dotfill	  	       &    -12.04 $_{    -0.23}^{     + 0.06}$ &    42.04 $_{	-0.24}^{   +   0.05}$	&     (f)   &CD 	    &	   AGN      &	 C	 \\
 IRAS\,17208-0014\dotfill    	       &    -13.45& 41.14	&     (e)   &CD 	    &	   AGN      &	 C	 \\
 NGC\,6482\dotfill	  	       &    -14.16 $_{    -0.30}^{     + 0.14}$ &    39.40 $_{	-0.37}^{   +   0.15}$	&     (f)   &SF+CD	    &	   SB	    &		\\
 NGC\,7130\dotfill	  	       &    -13.23 $_{    -0.11}^{     + 0.07}$ &    40.49 $_{	-0.13}^{   +   0.08}$	&     (f)   &		    &	   AGN      &	 C	 \\
 NGC\,7331\dotfill	  	       &    -13.73& 38.72	&     (e)   &		    &	   SB	    &	 C	\\
 IC\,1459\dotfill 	  	       &    -12.20 $_{    -0.02}^{     + 0.02}$ &    40.56 $_{	-0.03}^{   +   0.02}$		&     (f)   &SF 	    &	   AGN      &	 C	 \\
\hline
\end{tabular}
\end{center}
{\sc NOTES.-} Column (4): (f) Indicates flux and unabsorbed luminosity from spectral fitting and 
(e) indicates flux and luminosity estimated from empirical calibration. 
Column (5) gives those objects with obscuration indicators from spectral fitting (SF) 
of Color-color diagrams (CD); Column (6) gives our morphological classification 
attending to the 4.5--8.0 keV band; and column (7) gives the classification from the 
HST images, (C) Compact, (D) Dusty and (U) unclassified.
\end{table*}

\begin{table*}
\begin{center}
\caption{Results from spectral fitting and estimate from color-color 
diagrams.}\label{tab:fittings}
  \begin{tabular}{l@{~~~~~}c@{~~~~}c@{~~~~}c@{~~~~}c@{~~~}c@{~~}c@{~~}c@{~~}c@{~~}c@{~~}} \hline \hline \\ 
Name	  & Classif. & Model & $\rm{N_H}$ & $\rm{N_H(estim)}$ & $\rm{\Gamma}$ & $\rm{\Gamma}(estim)$& kT  & kT(estim) & $\rm{\chi^{2}}$/d.o.f\\
		&  &	& $(\rm{10^{22}cm^{-2}})$  & &	& &(keV) &(keV)&  \\  \hline \\
 NGC\,0315\dotfill	  &	AGN    &RS+PL&      0.71$_{-0.16}^{+0.07}$      &$\sim 10^{22}$&1.48$_{-0.19}^{+0.07}$      & [1.2-1.4] &  0.22$_{-0.01}^{ +0.08}$      & [0.7-0.8]&    188.77/161  \\
 ARP\,318A\dotfill	  &  	SB     & ... &       ...			&     ...      &...			    & [1.8-2.0]	&  ...  			& [0.6-0.8]&	...	    \\  		     
 ARP\,318B\dotfill	  &  	SB     & ... &       ...			&     ...      &...			    &  ...	&  ...  			& ...	   &	...	    \\  		     
 NGC\,1052\dotfill	  &  	AGN    & ... &       ...			&     ...      &...			    &  ...	&  ...  			& ...	   &	...	    \\  		     
 NGC\,2681\dotfill	  &	AGN    &ME+PL&      0.08$_{-0.08}^{+0.07}$      &     ...      &1.74$_{-0.47}^{+0.52}$      & [1.4-1.8] &  0.66$_{-0.08}^{ +0.09}$      & [0.7-0.8]&     13.94/26   \\ 
 UGC\,05101\dotfill	  &  	AGN    & ... &       ...			&     ...      &...			    & [0.4-0.6] &  ...  			& ...	   &	...	    \\  		     
 NGC\,3226\dotfill	  &  	AGN    & ... &       ...			&$\sim 10^{22}$&...			    & [0.8-1.2] &  ...  			& [0.7-0.8]&	...	    \\  		     
 NGC\,3245\dotfill	  &  	SB     & ... &       ...			&     ...      &...			    & [2.0-2.4] &  ...  			& [0.6-0.8]&	...	    \\  		     
 NGC\,3379\dotfill	  &	SB     & ... &       ...			&     ...      &...			    &  ...	&  ...  			& ...	   &	...	    \\ 
 NGC\,3507\dotfill	  &     SB     & ... &       ...			&     ...      &...			    &  ...	&  ...  			& ...	   &	...	    \\  		     
 NGC\,3607\dotfill	  &  	SB     & ... &       ...			&     ...      &...			    &  ...	&  ...  			& ...	   &	...	    \\  		     
 NGC\,3608\dotfill	  &  	SB     & ... &       ...			&     ...      &...			    &  ...	&  ...  			& [1.0-4.0]&	...	    \\  		     
 NGC\,3628\dotfill	  &  	AGN    & ... &       ...			&$\sim 10^{22}$&...			    & [1.2-1.6] &  ...  			& [0.6-0.8]&	...	    \\  		     
 NGC\,3690B\dotfill	  &	AGN    &PL   &      0.02$_{-0.02}^{+0.15}$      &     ...      &1.45$_{-0.25}^{+0.33}$      & [1.4-1.8] &  ...                          & [0.6-0.7]&     21.81/25   \\
 NGC\,4111\dotfill	  &  	AGN    & ... &       ...			&     ...      &...			    & [0.4-0.6] &  ...  			& [0.6-0.7]&	...	    \\  		     
 NGC\,4125\dotfill	  &  	AGN    & ... &       ...			&     ...      &...			    & [1.4-1.8] &  ...  			& [0.6-0.8]&	...	    \\  		     
 NGC\,4261\dotfill	  &  	AGN    & ... &       ...			&     ...      &...			    & [1.5-1.6] &  ...  			& ...	   &	...	    \\  		     
 NGC\,4314\dotfill	  &  	SB     & ... &       ...			&     ...      &...			    &  ...	&  ...  			& ...	   &	...	    \\  		     
 NGC\,4374\dotfill	  &	AGN    &PL   &      0.14$_{-0.05}^{+0.06}$      &    ...       &2.07$_{-0.20}^{+0.24}$      & [2.0-2.2] &  ...                          & [0.7-0.8]&     32.20/33   \\
 NGC\,4395\dotfill	  &     AGN    &PL   &      2.87$_{-0.38}^{+0.53}$      &    ...       &1.44$_{-0.19}^{+0.26}$      &  ...	&  ...                          & ...	   &    123.68/106  \\
 NGC\,4410A\dotfill	  &     AGN    & PL  &      0.01$_{-0.01}^{+0.04}$      &    ...       &1.75$_{-0.12}^{+0.19}$      & [1.4-1.8] &  ...                          & [0.7-0.8]&     33.11/45   \\
 NGC\,4438\dotfill	  &     SB     &RS+PL&      0.11$_{-0.04}^{+0.07}$      &    ...       &2.13$_{-0.62}^{+0.67}$      & [1.8-2.0] &  0.78$_{-0.06}^{ +0.04}$      & [0.7-0.9]&     32.27/34   \\				  
 NGC\,4457\dotfill	  &     AGN    &RS+PL&      0.07$_{-0.07}^{+0.08}$      &    ...       &1.98$_{-0.54}^{+0.66}$      & [1.6-1.8] &  0.66$_{-0.08}^{ +0.11}$      & [0.6-0.8]&     18.68/18   \\
 NGC\,4459\dotfill	  &  	SB     & ... &       ...			&     ...      &...			    &   ...	&  ...  			& ...	   &	...	    \\  		     
 NGC\,4486\dotfill	  &  	AGN    & ... &       ...			&     ...      &...			    & [1.2-1.6] &  ...  			& [0.7-0.8]&	...	    \\  		     
 NGC\,4494\dotfill	  &     AGN    &ME+PL&      0.11$_{-0.01}^{+0.15}$      &    ...       &1.37$_{-0.88}^{+0.91}$      &  ...	&  0.63$_{-0.13}^{ +0.20}$      & ...	   &      5.59/5	 \\ 
 NGC\,4552\dotfill	  &     AGN    &RS+PL&      0.04$_{-0.04}^{+0.02}$      &    ...       &1.81$_{-0.10}^{+0.24}$      & [1.8-2.0] &  0.83$_{-0.07}^{ +0.05}$      & [0.7-0.9]&     57.19/61   \\
 NGC\,4579\dotfill	  &  	AGN    & ... &       ...			&     ...      &...			    & [1.2-1.6] &  ...  			& ...	   &	...	    \\  		     
 NGC\,4594\dotfill~.~.~.~.~.~.~ &AGN   &PL   &      0.20$_{-0.04}^{+0.02}$      &$\sim 10^{22}$&1.41$_{-0.10}^{+0.11}$      & [1.6-2.0] &  ...                          & [0.4-0.6]&    118.72/87   \\
 NGC\,4596\dotfill	  &  	 SB    & ... &       ...			&     ...      &...			    &  ...	&  ...  			& ...	   &	...	    \\  		     
 NGC\,4636\dotfill	  &  	 SB    & ... &       ...			&     ...      &...			    & [2.0-2.2] &  ...  			& [0.7-0.8]&	...	    \\  		     
 NGC\,4676A\dotfill	  &  	 SB    & ... &       ...			&     ...      &...			    & [0.8-1.2] &  ...  			& ...	   &	...	    \\  		     
 NGC\,4676B\dotfill	  &  	 SB    & ... &       ...			&     ...      &...			    &  ...  	&  ...  			& ...	   &	...	    \\  		     
 NGC\,4696\dotfill	  &	 SB    &PL   &      0.13$_{-0.13}^{+0.18}$      &    ...       &4.26$_{-0.62}^{+0.96}$      &  ...	&  ...                          &  ...	   &     21.98/17   \\
 NGC\,4698\dotfill	  &  	 SB    & ... &       ...			&     ...      &...			    & [1.2-1.6] &  ...  			& [0.7-0.8]&	...	    \\  		     
 NGC\,4736\dotfill	  &	 AGN   &ME+PL&      0.08$_{-0.04}^{+0.04}$      &    ...       &2.00$_{-0.06}^{+0.23}$      & [1.4-1.8] &  0.60$_{-0.09}^{ +0.05}$      & [0.6-0.8]&     72.90/79   \\ 
 NGC\,5055\dotfill	  &  	 AGN   & ... &       ...			&     ...      &...			    & [1.6-2.0] &  ...  			& [0.7-0.9]&	...	    \\  		     
 NGC\,5194\dotfill        &	 AGN   &ME+PL&      0.03$_{-0.03}^{+0.06}$      &    ...       &2.67$_{-0.47}^{+0.71}$      & [0.4-0.6] &  0.61$_{-0.04}^{ +0.03}$      & [0.6-0.7]&     49.74/48   \\ 
 MRK\,266NE\dotfill	  &  	AGN    & ... &       ...			&     ...      &...			    &  ...	&  ...  			&  ...	   &	...	    \\  		     
 UGC\,08696\dotfill	  &	 AGN   &RS+PL&      0.39$_{-0.06}^{+0.06}$      &    ...       &1.74$_{-0.61}^{+0.81}$      &  ...	&  0.75$_{-0.12}^{ +0.12}$      & ...	   &     55.63/40   \\
 CGCG\,162-010\dotfill	  &      SB    &RS+PL&      0.09$_{-0.09}^{+0.21}$      &    ...       &2.44$_{-0.43}^{+1.51}$      &  ...	&  1.12$_{-0.21}^{ +0.18}$      & ...	   &     59.95/61   \\
 NGC\,5746\dotfill	  &	 AGN   &PL   &      0.60$_{-0.52}^{+0.36}$      &$\sim 10^{22}$&1.22$_{-0.39}^{+0.14}$      & [1.0-1.2] &  ...                          & [0.7-0.8]&     15.66/17   \\
 NGC\,5846\dotfill	  &	 SB    &ME+PL&      0.12$_{-0.12}^{+0.35}$      &    ...       &1.65$_{-0.67}^{+0.95}$      & [0.8-1.2] &  0.50$_{-0.21}^{ +0.11}$      & [0.5-0.7]&     21.91/25   \\ 
 NGC\,5866\dotfill	  &  	 SB    & ... &       ...			&     ...      &...			    &  ...	&  ...  			& ...	   &	...	    \\  		     
 NGC\,6251\dotfill	  &	 AGN   &ME+PL&      0.44$_{-0.26}^{+0.43}$      &$\sim 10^{22}$&1.60$_{-0.22}^{+0.29}$      & [1.0-1.4] &  0.26$_{-0.09}^{ +0.34}$      & [0.4-0.6]&     65.27/62   \\ 
 NGC\,6240\dotfill	  &	 AGN   &RS+PL&      1.11$_{-0.10}^{+0.10}$      &$\sim 10^{22}$&1.03$_{-0.15}^{+0.14}$      & [0.8-1.0] &  0.76$_{-0.06}^{ +0.06}$      & [0.7-0.8]&    126.79/112  \\
 IRAS\,17208-0014\dotfill &  	AGN    & ... &       ...			&$\sim 10^{22}$&...			    & [1.6-2.0] &  ...  			& [0.6-0.8]&	...	    \\  		     
 NGC\,6482\dotfill	  &	 SB    &ME   &      0.31$_{-0.18}^{+0.10}$      &    ...       &...                         &   ...	&  0.68$_{-0.07}^{ +0.13}$      & ...	   &     16.41/18   \\
 NGC\,7130\dotfill	  &	 AGN   &RS+PL&      0.06$_{-0.04}^{+0.04}$      &    ...       &2.51$_{-0.22}^{+0.41}$      & [1.2-1.4] &  0.82$_{-0.05}^{ +0.03}$      & [0.7-0.8]&     81.43/57   \\ 
 NGC\,7331\dotfill	  &  	 SB    & ... &       ...			&     ...      &...			    & [2.2-2.6] &  ...  			& [0.6-0.8]&	...	    \\  		     
IC\,1459\dotfill 	  &      AGN   &ME+PL&      0.28$_{-0.02}^{+0.15}$      &    ...       &1.89$_{-0.11}^{+0.10}$      & [1.8-2.0] &  0.30$_{-0.06}^{ +0.25}$      & [0.7-0.9]&    189.47/166  \\ \hline
\end{tabular}
\end{center}
{\sc NOTES.-} (*) Out of any grid. UGC\,08696 has been fitted including a gaussian model where the best fit is centered in 5.93 keV with
a weight of 1.22~keV.
\end{table*}

\begin{table*}[!h]
\begin{center}
 \caption{Soft ($Q_{A}$), medium ($Q_{B}$) and hard ($Q_{C}$) hardness ratios for the
 whole sample.}\label{tab:counts}
\begin{tabular}{@{~~~~~~~~~~~~~}c@{~~~~~~~~~~~}r@{~}l@{~~~~~~~~~~}r@{~}l@{~~~~~~~~~~}r@{~}l@{~~~~~~~~~~~~~}} \hline \hline \\
     Name      &  $Q_{A}$     &  &  $Q_{B}$      & & $Q_{C}$ &       \\ \hline
NGC\,0315\dotfill       &     -0.06   & $\pm$0.05  &	-0.05	&   $\pm$0.04 &     -0.59 &	$\pm$0.03    \\
ARP\,318A\dotfill       &     -0.20   & $\pm$0.11  &	-0.45	&   $\pm$0.14 &     -0.74 &	$\pm$0.17    \\
ARP\,318B\dotfill       &      0.04   & $\pm$0.18  &	-0.09	&   $\pm$0.22 &     -0.88 &	$\pm$0.10    \\
NGC\,1052\dotfill       &     -0.13   & $\pm$0.21  &	 0.03	&   $\pm$0.23 &     -0.37 &	$\pm$0.13    \\
NGC\,2681\dotfill       &     -0.09   & $\pm$0.06  &	-0.27	&   $\pm$0.11 &     -0.67 &	$\pm$0.13    \\
UGC\,05101\dotfill      &     -0.03   & $\pm$0.13  &	-0.05	&   $\pm$0.14 &     -0.41 &	$\pm$0.14    \\
NGC\,3226\dotfill       &      0.58   & $\pm$0.13  &	 0.09	&   $\pm$0.14 &     -0.49 &	$\pm$0.08    \\
NGC\,3245\dotfill       &     -0.19   & $\pm$0.26  &	-0.35	&   $\pm$0.29 &     -0.78 &	$\pm$0.20    \\
NGC\,3379\dotfill       &      0.18   & $\pm$0.22  &	-0.34	&   $\pm$0.28 &      ...  &	 ...         \\
NGC\,3507\dotfill       &     -0.36   & $\pm$0.08  &	-0.40	&   $\pm$0.20 &      ...  &	 ...	     \\
NGC\,3607\dotfill       &      0.25   & $\pm$0.17  &	-0.38	&   $\pm$0.24 &      ...  &	 ...	     \\
NGC\,3608\dotfill       &      0.47   & $\pm$0.22  &	-0.69	&   $\pm$0.17 &     -0.75 &	$\pm$0.23    \\
NGC\,3628\dotfill       &      0.46   & $\pm$0.26  &	 0.10	&   $\pm$0.17 &     -0.60 &	$\pm$0.17    \\
NGC\,3690B\dotfill      &      0.39   & $\pm$0.12  &	-0.25	&   $\pm$0.09 &     -0.48 &	$\pm$0.10    \\
NGC\,4111\dotfill       &     -0.26   & $\pm$0.08  &	-0.38	&   $\pm$0.18 &     -0.45 &	$\pm$0.17    \\
NGC\,4125\dotfill       &     -0.18   & $\pm$0.16  &	-0.32	&   $\pm$0.19 &     -0.68 &	$\pm$0.21    \\
NGC\,4261\dotfill       &     -0.37   & $\pm$0.03  &	-0.36	&   $\pm$0.07 &     -0.33 &	$\pm$0.06    \\
NGC\,4314\dotfill       &     -0.22   & $\pm$0.16  &	-0.35	&   $\pm$0.29 &      ...  &      ...         \\
NGC\,4374\dotfill       &     -0.07   & $\pm$0.08  &	-0.22	&   $\pm$0.09 &     -0.75 &	$\pm$0.08    \\
NGC\,4395\dotfill       &     -0.20   & $\pm$0.12  &	 0.34	&   $\pm$0.06 &     -0.52 &	$\pm$0.03    \\
NGC\,4410A\dotfill      &     -0.07   & $\pm$0.06  &	-0.27	&   $\pm$0.08 &     -0.67 &	$\pm$0.07    \\
NGC\,4438\dotfill       &     -0.05   & $\pm$0.05  &	-0.41	&   $\pm$0.09 &     -0.75 &	$\pm$0.13    \\
NGC\,4457\dotfill       &     -0.17   & $\pm$0.07  &	-0.25	&   $\pm$0.12 &     -0.70 &	$\pm$0.14    \\
NGC\,4459\dotfill       &     -0.13   & $\pm$0.25  &	 0.00	&   $\pm$0.43 &      ...  &      ...         \\
NGC\,4486\dotfill       &     -0.04   & $\pm$0.01  &	-0.21	&   $\pm$0.02 &     -0.64 &	$\pm$0.01    \\
NGC\,4494\dotfill       &     -0.02   & $\pm$0.12  &	-0.16	&   $\pm$0.16 &     -0.87 &	$\pm$0.08    \\
NGC\,4552\dotfill       &     -0.01   & $\pm$0.05  &	-0.20	&   $\pm$0.07 &     -0.69 &	$\pm$0.07    \\
NGC\,4579\dotfill       &     -0.08   & $\pm$0.02  &	-0.15	&   $\pm$0.02 &     -0.61 &	$\pm$0.02    \\
NGC\,4594\dotfill       &      0.15   & $\pm$0.05  &	-0.14	&   $\pm$0.05 &     -0.65 &	$\pm$0.04    \\
NGC\,4596\dotfill       &     -0.42   & $\pm$0.37  &	-0.35	&   $\pm$0.56 &      ...  &      ...         \\
NGC\,4636\dotfill       &     -0.12   & $\pm$0.03  &	-0.50	&   $\pm$0.06 &     -0.73 &	$\pm$0.11    \\
NGC\,4676A\dotfill      &      0.07   & $\pm$0.25  &	-0.10	&   $\pm$0.29 &     -0.53 &	$\pm$0.40    \\
NGC\,4676B\dotfill      &     -0.21   & $\pm$0.18  &	-0.16	&   $\pm$0.25 &      ...  &      ...         \\
NGC\,4696\dotfill       &     -0.17   & $\pm$0.11  &	 0.00	&   $\pm$0.33 &      ...  &      ...         \\
NGC\,4698\dotfill       &     -0.06   & $\pm$0.17  &	-0.20	&   $\pm$0.22 &     -0.63 &	$\pm$0.25    \\
NGC\,4736\dotfill       &     -0.15   & $\pm$0.04  &	-0.23	&   $\pm$0.05 &     -0.68 &	$\pm$0.06    \\
NGC\,5055\dotfill       &      0.03   & $\pm$0.14  &	-0.28	&   $\pm$0.15 &     -0.73 &	$\pm$0.13    \\
NGC\,5194\dotfill       &     -0.25   & $\pm$0.03  &	-0.42	&   $\pm$0.07 &     -0.45 &	$\pm$0.17    \\
MRK\,0266NE\dotfill     &     -0.01   & $\pm$0.13  &	 0.13	&   $\pm$0.19 &     -0.34 &	$\pm$0.14    \\
UGC\,08696\dotfill      &      0.00   & $\pm$0.10  &	-0.14	&   $\pm$0.11 &      0.12 &	$\pm$0.08    \\
CGCG\,162-010.~.~.~.\dotfill&  0.09   & $\pm$0.07  &	-0.22	&   $\pm$0.12 &      ...  &      ...         \\
NGC\,5746\dotfill       &      0.58   & $\pm$0.18  &	-0.01	&   $\pm$0.12 &     -0.51 &	$\pm$0.08    \\
NGC\,5846\dotfill       &     -0.35   & $\pm$0.05  &	-0.35	&   $\pm$0.12 &     -0.57 &	$\pm$0.26    \\
NGC\,5866\dotfill       &     -0.23   & $\pm$0.21  &	-0.63	&   $\pm$0.29 &     -0.52 &	$\pm$0.24    \\
NGC\,6251\dotfill       &      0.19   & $\pm$0.07  &	-0.21	&   $\pm$0.06 &     -0.55 &	$\pm$0.05    \\
NGC\,6240\dotfill       &      0.52   & $\pm$0.05  &	 0.01	&   $\pm$0.05 &     -0.48 &	$\pm$0.05    \\
IRAS\,17208-0014\dotfill &      0.44   & $\pm$0.18  &	 0.01	&   $\pm$0.15 &     -0.67 &	$\pm$0.12    \\
NGC\,6482\dotfill       &      0.10   & $\pm$0.08  &	-0.38	&   $\pm$0.14 &     -0.83 &	$\pm$0.15    \\
NGC\,7130\dotfill       &     -0.12   & $\pm$0.04  &	-0.33	&   $\pm$0.06 &     -0.61 &	$\pm$0.09    \\
NGC\,7331\dotfill       &     -0.16   & $\pm$0.21  &	-0.35	&   $\pm$0.22 &     -0.81 &	$\pm$0.14    \\
IC\,1459\dotfill        &      0.00   & $\pm$0.03  &	-0.19	&   $\pm$0.03 &     -0.72 &	$\pm$0.03    \\ \hline
\end{tabular}
\end{center}
\end{table*}
\twocolumn

\end{document}